\documentstyle[aps,multicol,prb,epsf]{revtex}
\begin{document}
\draft
\title{Effects of external radiation on biased Aharonov-Bohm rings}
%\title{Persistent Currents in Open Rings}
\author{O. Entin-Wohlman$^{a,b,c,}$\thanks{On leave from
the School of Physics and Astronomy, Raymond and Beverly Sackler
Faculty of Exact Sciences,  Tel Aviv University, Tel Aviv 69978,
Israel.},  Y. Imry$^d$, and A. Aharony$^{a,b,*}$}
\address{$^{a}$Materials Science Division, Argonne National
Laboratory, Argonne, Illinois 60439, USA}
%\address{$^a$School of Physics and Astronomy, Raymond and Beverly Sackler
%Faculty of Exact Sciences, \\ Tel Aviv University, Tel Aviv 69978,
%Israel\\ }
\address{$^b$Department of Physics, Ben Gurion University, Beer
Sheva 84105, Israel}
\address{$^c$Albert Einstein
Minerva Center for Theoretical Physics \\ at the Weizmann
Institute of Science,  Rehovot 76100, Israel}
\address{$^d$Department of Condensed Matter Physics, The Weizmann
Institute of Science, Rehovot 76100, Israel}
\date{\today}
\maketitle
%first comas and then references

\begin{abstract}
We consider  the currents flowing in a solid-state interferometer
under the effect of both an Aharonov-Bohm phase and a bias
potential. Expressions are obtained for these currents, allowing
for electronic or electron-boson interactions, which may take
place solely on a quantum dot placed on one of the interferometer
arms. The boson system can be out of  equilibrium.
%, and may induce
%a nonequilibrium electronic distribution.
The results are used to obtain the transport current through the
interferometer, and the current circulating around it under the
effect of the Aharonov-Bohm flux. The modifications of both
currents, brought about by coupling the quantum dot to an
incoherent sonic or electromagnetic source, are then analyzed. By
choosing the appropriate range of the boson source intensity and
its frequency, the magnitude of the interference-related terms of
both currents can be controlled.
\end{abstract}


\begin{multicols}{2}

\section{Introduction}

Solid-state interferometers, restricted to the mesoscopic scale in
order to retain the coherence of conduction electrons, \cite{joe}
are constructed from narrow waveguides, possibly containing
scatterers, for the electronic paths. An Aharonov-Bohm magnetic
flux \cite{AB} between the two paths in such interferometers
results in a periodic flux-dependence behavior, which stems from
interference of the electronic wave-functions. In recent
experiments,
\cite{amir,schuster,sprinzak,kouwenhoven,ji,holleitner,fuhrer,kobayashi}
carried on  interferometers connected to several electronic
reservoirs, the current passing through the system in response to
a voltage difference has been used to investigate {\it coherent
transport}. These experiments have revived interest in such
systems, whose theoretical \cite{gefen,pinhas,claudio} and
experimental \cite{webb} study has begun much earlier.  The
current experimental set-ups involve a quantum dot (or two
\cite{holleitner,fuhrer}) embedded in the interferometer, aiming
at the study of the transmission properties of the former. These
experiments have been followed by many theoretical works,
exploring the possibility of deducing the transmission phase of a
quantum dot from the measured conductance of the interferometer,
\cite{hackenbroich,bruder,lee,taniguchi,yeyati,gerland,bulka,hofstetter,konig,avi,weidenmuller}
and investigating its dependence on various interactions.

The interference of the electronic wave functions in an
Aharonov-Bohm interferometer also creates a circulating current,
which flows even at thermal equilibrium, and even when the ring is
isolated (under these conditions it is usually called `persistent
current'). This current has been invoked as early as 1936 by
Pauling, \cite{Pauling}  to explain the large orbital magnetic
response of $\pi$ electrons moving on a ring in benzene-type
molecules, and soon after has been calculated \cite{london} in
terms of the tight-binding model. The analogy between persistent
currents and the Josephson effect has been expounded upon in Refs.
\onlinecite{kulik} and \onlinecite{buttiker}. Their discussion of
the possible realization of a `normal Josephson current' in small
metallic (or semiconductor) rings,  in the presence of some
disorder, has sparked much interest in this phenomenon and led to
a considerable experimental effort to detect it, either by various
magnetic response measurements,
\cite{levy,chandrasekhar,mailly,jariwala,rabaud,bouchiat} or by
optical spectroscopy. \cite{lorke,warburton,bayer} At thermal
equilibrium, the  persistent current is equivalent to the
thermodynamic orbital magnetic moment of the electrons. Since it
arises from the interference of the electronic wave functions,
then, as long as the electrons are phase-coherent, it will survive
the presence of moderate static disorder. \cite{joe,buttiker}
Recently, most of the theoretical interest in this phenomenon has
shifted to studying charge- (or spin) fluctuation effects,
\cite{staford,cho,ding}  time-dependent properties and
non-equilibrium situations,
\cite{kravtsov,galperin,moskalets,chalaev} or electronic
interactions. \cite{ferrari,kang,affleck,hu,eckle} In addition,
there have been recently several attempts
\cite{mohanty,schwab,altshuler,cedrashi} to relate the phenomenon
of persistent current, which is intimately connected to electronic
coherence,  to the dephasing of electrons at equilibrium due to
the coupling with a boson bath.

\begin{figure}[htb]
\leavevmode \epsfclipon \epsfxsize=7.truecm
\vbox{\epsfbox{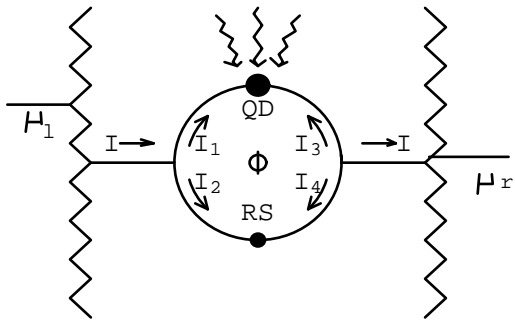}} \vspace{0.5cm}
 \caption{An Aharonov-Bohm interferometer,
containing a quantum dot (QD) on its upper arm and threaded by a
magnetic flux $\Phi$.  The lower arm of the interferometer
contains a `reference' site (RS). The ring is connected to two
electronic reservoirs whose chemical potentials are either equal
or have a small difference, allowing a current $I$ to flow from
the left to the right. The wavy vectors denote the external beam
radiated on the dot.} \label{fig1}
\end{figure}

Here we study the currents flowing {\it around and through} an
`open' interferometer, connected to electronic reservoirs, with a
quantum dot placed on one of its arms, when the latter is coupled
to an external incoherent radiation source. The electronic
reservoirs are held at slightly different chemical potentials,
such that the voltages are  small enough for the system to be in
the {\it linear transport} regime. The external radiation source,
on the other hand, will be taken as being, and possibly driving
the system, out of equilibrium, so that its intensity can serve as
a `control parameter' of the currents. In other words, we study
the currents when the electrons are also coupled to an incoherent
{\it out-of-equilibrium} boson source. We take the electronic
system to be free of any interactions, except on the quantum dot,
where the electrons are coupled to an external source of sonic (or
electromagnetic) waves. The system is depicted in Fig. \ref{fig1}.

Although we  use the term `electron-phonon interaction' throughout
this paper, our results apply equally, with minor modifications,
to the case where the electrons are coupled  to an
electro-magnetic source, that is, for the electron-photon
interaction. In any event, in order to retain the coherence of the
electrons, the systems we consider are necessarily confined to
size scales small enough so that the electrons stay coherent at
the given temperature. At the same time, the strength of the
acoustic source is assumed to be such that the additional
decoherence due to it is not detrimental. The precise parameter
windows in which this can be achieved is sensitive to acoustic (or
electromagnetic) mismatch, details of the sample geometries, etc,
and hence their calculations are not  carried out here. Also, we
do not discuss dephasing, but rather, like in Refs.
~\onlinecite{kravtsov} and \onlinecite{chalaev}, we concentrate on
a {\it non-equilibrium} source of bosons.

\vspace{0.5cm}

\vspace{0.5cm}

When the electrons are coupled to a boson source, the na\"{\i}ve
expectation is that  the coherent current decreases due to loss of
coherence, caused by inelastic processes as well as by
renormalization effects due to the `dressing' of the electrons by
the bosons (the polaron effect \cite{holstein}). The latter is
manifested by an overall Debye-Waller exponent. However, it turns
out that this is not the whole effect brought about by the
radiation. In the case of {\it isolated} rings, it has been found
\cite{aronov} that when the electrons are coupled to phonons, the
persistent current is not only diminished; Rather, there appears
an additional term, which originates from delicate resonance
processes in which at least two phonons are involved (those were
termed `doubly-resonant processes'). The additional orbital
magnetic moment appears at non-zero temperatures, and has a
non-monotonic temperature dependence at sufficiently low
temperatures. \cite{aronov} At thermal equilibrium, this new term
has been found to further reduce the persistent current (beyond
its value in the absence of the coupling to the bosons). However,
at non-equilibrium situations, the magnitude of that `extra'
contribution may be tuned by controlling the intensity of the
radiation  in a certain frequency range, which is experimentally
feasible. Possibly related experiments with extremely interesting
results have recently been reported in, e.g., Refs.
\onlinecite{cavill} and \onlinecite{mani}. Here we examine the
effect of the electron-phonon coupling on a biased Aharonov-Bohm
interferometer, which consists of an `open' ring, connected to two
reservoirs. Then, in addition to the circulating current induced
by the magnetic phase, there appears also a transport current. We
find that in a certain sense, the open ring is more amenable to
manipulations by an external radiation source. We show that both
the circulating and the transport  currents are affected by a
radiation source in a similar manner: Beside  the overall
Debye-Waller factor, they each acquire an additional contribution.
In the case of an open ring, that additional term does not
necessitate the existence of real resonant transitions between the
initial and the final state, it appears at a lower order in the
electron-phonon interaction, (as compared to the situation in
isolated rings) and it exists even at zero temperature. The
magnitude of that contribution can again be tuned by controlling
the intensity of the radiation in a certain frequency range. In
other words, by coupling the electrons to an {\it
out-of-equilibrium} radiation source, one may control both the
circulating and the transport current. Such a relation between the
radiation intensity and the orbital magnetic moment may open
interesting possibilities for future nano-devices.

Our method of calculation is to express all partial currents
flowing in the system (i.e., $I_{1}$, $I_{2}$, $I_{3}$, and
$I_{4}$, see Fig. \ref{fig1}) in terms of the exact (and
generally, unknown) Green function on the dot, which includes all
the effects of the coupling to the interferometer, the external
reservoirs, and the interactions taking place on the dot. These
expressions do not necessitate a near-equilibrium situation. In so
doing, we derive general expressions for the current passing
through the interferometer, $I$, and the current circulating
around it which is induced by the Aharonov-Bohm flux, in terms of
the exact Green function on the dot. We then use these results to
investigate the effect of coupling to a boson source on both
currents.

We begin in Sec. II by the derivation of the partial currents, the
transport current, and the circulating current. The expressions we
obtain are valid also for the case in which the electrons
experience electronic interactions on the dot. In particular, our
result for the transport current generalizes  the ones reported
previously, \cite{bulka,hofstetter}  which were derived under the
assumption that there is no scattering on the reference arm.
Within that approximation, the flux-dependence of the linewidth on
the quantum dot level is lost. This flux dependence, as we  show,
turns out to be crucial in determining the circulating current.
Section  II is supplemented by an Appendix, detailing the
computation of the partial currents (Appendix A). In sec. III we
employ the general result for the transport current to study the
effect of the coupling to a boson source. To this end, we use an
approximate expression for that dot Green function,
\cite{glazman,wingreen,jauho} for the case in which the electrons
on the dot are coupled linearly to a sonic source. Section IV is
devoted to the analysis of the  circulating current under
irradiation. For the sake of completeness, we include in that
section a discussion of the effect of electron-phonon coupling on
electrons moving on electronically-isolated rings (which are
decoupled from the leads). Finally, we summarize our findings in
Sec. V.

\section{The currents in a biased interferometer}

Figure \ref{fig1} portrays an Aharonov-Bohm interferometer, with a
quantum dot placed on its upper arm, and a second electronic site
placed on the other arm, serving as a `reference' site. All
interactions (among the electrons, or electron-boson interactions)
are taking place only on the quantum dot. The interferometer is
connected at the left and at the right to electronic  reservoirs,
kept at slightly different  (or equal) chemical potentials. The
connection is via  single-channel leads. The model Hamiltonian
describing this system consists of four terms,
\begin{eqnarray}
{\cal H}={\cal H}_{\rm leads}+{\cal H}_{\rm ref}+{\cal H}_{\rm
d}+{\cal H}_{\rm tun},\label{ham}
\end{eqnarray}
in which the first  term describes the leads, which are assumed to
be two free-electron systems,
\begin{eqnarray}
{\cal H}_{\rm
leads}=\sum_{k}\epsilon_{k}c^{\dagger}_{k}c_{k}+\sum_{p}\epsilon_{p}c^{\dagger}_{p}c_{p}.
\end{eqnarray}
(We  omit spin indices when they are not necessary.) The left lead
states are denoted by $k$, and the right ones by $p$, with $c_{k}$
($c_{p}$) being the destruction operator for states on the left
(right) lead. For one-dimensional leads, described by a
tight-binding model with a nearest-neighbor hopping matrix element
$J$, one has $\epsilon_{k}=-2J\cos k$, and similarly
$\epsilon_{p}=-2J\cos p$. The chemical potential in the reservoir
connected to the left lead, $\mu_{\ell}$, can differ from that on
the right reservoir, $\mu_{r}$. Otherwise, the two leads are taken
as identical, i.e., they have the same large bandwidth $2J$. The
reference site is taken for simplicity as having a single
localized level, of energy $\epsilon_{0}$; Hence
\begin{eqnarray}
{\cal H}_{\rm ref}=\epsilon_{0}c^{\dagger}_{0}c_{0}.
\end{eqnarray}
The dot Hamiltonian ${\cal H}_{\rm d}$ is not  specified at the
moment; It may include electron-electron interactions or
electron-phonon interactions. For simplicity, we assume that only
one of the dot single-energy levels is effectively connected to
the leads. It is possible to carry out a more general calculation;
However, the algebra then becomes complicated and may obscure the
physical effects we wish to explore. Hence we write for the
tunneling Hamiltonian
\begin{eqnarray}
{\cal H}_{\rm
tun}&=&\sum_{k}V_{k}c^{\dagger}_{k}d+\sum_{p}V_{p}c^{\dagger}_{p}d\nonumber\\
&+&\sum_{k}\upsilon_{k}c^{\dagger}_{k}c_{0}+\sum_{p}\upsilon_{p}c^{\dagger}_{p}c_{0}+hc,
\end{eqnarray}
where $d$ is the destruction operator for the electron on the dot.
The tunneling matrix elements for a one-dimensional tight-binding
model read
\begin{eqnarray}
&&V_{k}=-\sqrt{\frac{2}{N}}j_{\ell }\sin k ,\ \
V_{p}=-\sqrt{\frac{2}{N}}j_{r }\sin p,\nonumber\\
&&\upsilon_{k}=-\sqrt{\frac{2}{N}}i_{\ell}e^{i\phi_{\ell}}\sin k,\
 \upsilon_{p}=-\sqrt{\frac{2}{N}}i_{r}e^{-i\phi_{r}}\sin
p,\label{coupling}
\end{eqnarray}
where $N$ is the number of sites on each of the leads, and gauge
invariance allows one to assign the flux dependence to the
reference arm, such that the total flux (which is the magnetic
flux threading the ring, measured in units of the flux quantum) is
\begin{eqnarray}
\Phi=\phi_{\ell}+\phi_{r}.
\end{eqnarray}
In Eq. (\ref{coupling}), $j_{\ell}$ and $j_{r}$ are the matrix
elements coupling the dot to the left and right point contacts,
and $i_{\ell}$ and $i_{r}$ are those connecting the reference site
to the same points. We emphasize that the model considered here
does {\em not allow \cite {avi} for any electron losses}. This is
often referred to as a ``closed interferometer".

Under the circumstances described above, a transport current $I$
is passed through the ring, say from left to right. This current
splits into the currents moving in the upper and  lower arms of
the ring,
\begin{eqnarray}
I=I_{1}+I_{2}.\label{Itran}
\end{eqnarray}
When all electrons entering the interferometer from the left
reservoir leave it into the right one, and are not lost to the
surrounding, (as sometimes happens in the experiments), one has
$I_{1}+I_{3}=I_{2}+I_{4}=0$. For  reasons related to the detailed
calculations below, we keep the four partial currents separately.
The  current {\it circulating} the ring under the effect of the
Aharonov-Bohm flux, $I_{cir}$, is conveniently defined as
\begin{eqnarray}
2I_{cir}&=&\frac{1}{2}(I_{1}-I_{2})\Bigl |_{\Phi}-
\frac{1}{2}(I_{1}-I_{2})\Bigl |_{-\Phi},\label{Ipc}
\end{eqnarray}
in order to avoid spurious currents caused by geometrical
asymmetries. It is therefore seen that the calculation of both the
transport current and the circulating one  requires the knowledge
of the partial currents in the interferometer.

An efficient way to find those currents is to employ the Keldysh
technique, which is particularly suitable to handle
non-equilibrium situations. \cite{langreth}  Using the Keldysh
notations, the partial currents $I_{1}$ and $I_{2}$ are given by
\cite{jauho} (in units in which $\hbar=1$ )
\begin{eqnarray}
I_{1}&=&e\int\frac{d\omega}{2\pi}\sum_{k}V_{k}\Bigl
(G^{<}_{kd}(\omega )-G^{<}_{dk}(\omega )\Bigr
),\nonumber\\
I_{2}&=&e\int\frac{d\omega}{2\pi}\sum_{k}\Bigl
(\upsilon^{\ast}_{k} G^{<}_{k0}(\omega
)-\upsilon_{k}G^{<}_{0k}(\omega )\Bigr ), \label{currents}
\end{eqnarray}
where
\begin{eqnarray}
G^{<}_{ab}(\omega )&=&\int dt e^{i\omega t}i\langle
b^{\dagger}a(t)\rangle,\label{KELDYSH}
\end{eqnarray}
and the operators $a$ and $b$ stand for $c_{k}$, $c_{p}$, $c_{0}$,
or $d$. The other two partial currents, $I_{3}$ and $I_{4}$, are
derived from Eqs. (\ref{currents}) by changing the lead index $k$
into the second lead index, $p$.

The computation of all four partial currents is detailed in
Appendix A. Here we summarize the results. The first step taken
there is to obtain explicit expressions  [see Eqs. (\ref{I1}) and
(\ref{I2})] for the partial currents in terms of the various
parameters, and the {\it exact} Green function on the dot, which
includes all effects of interactions, as well as the couplings to
the interferometer, to the electronic reservoirs and to the phonon
source. In the Keldysh technique this means that the
above-mentioned expressions include the Keldysh function
$G_{dd}^{<}$, [see Eq. (\ref{KELDYSH})], and the usual retarded
($G^{R}_{dd}$) and the advanced [$G^{A}_{dd}=(G^{R}_{dd})^{\ast}$]
dot Green functions. The frequency ($\omega$)- integration of the
former, $\int d\omega G^{<}_{dd}$, has a very clear physical
meaning: It gives the occupation number of the electrons on the
dot, $n_{d}$.

When the interferometer is biased, the Keldysh Green function
$G^{<}_{dd}$ and the occupation $n_{d}$ are affected by the
voltage difference, such that current conservation,
$I_{1}+I_{3}=0$, is ensured (see Fig. \ref{fig1}). In practice,
however, the calculation of the Keldysh function is not simple
(except for the interaction-free system). We therefore resort to
an approximation, which gives it in terms of $G_{dd}^{R}$ and
$G_{dd}^{A}$. Explicitly, one finds (see Appendix A2 for details)
\begin{eqnarray}
I_{1}+I_{3}&=&e\int\frac{d\omega}{2\pi}\Bigl [(\Sigma^{R}_{\rm
ext}-\Sigma^{A}_{\rm ext})G^{<}_{dd}\nonumber\\
&+&\Sigma_{\rm ext}^{<}(G^{A}_{dd}-G^{R}_{dd})\Bigr ],\label{cons}
\end{eqnarray}
where the frequency dependence of the various functions is
suppressed for brevity. Here, $\Sigma_{\rm ext}^{R}$ is that part
of the retarded self-energy on the dot, which comes solely from
the couplings to the interferometer and to the leads. Namely, it
is the self-energy part for the interaction-free system.
Similarly, $\Sigma^{A}_{\rm ext}=(\Sigma^{R}_{\rm ext})^{\ast}$ is
the advanced self-energy coming from those couplings, and
$\Sigma^{<}_{\rm ext}$ is the corresponding Keldysh function. All
the above three self-energies can be found quite
straightforwardly, as they pertain to the non-interacting parts of
the Hamiltonian [see Eqs. (\ref{sigmaext}) and (\ref{sigmaextK})]
. When the system is free of interactions, or when it is
un-biased, namely, $\mu_{\ell}=\mu_{r}$ (see Appendix A2), the
integrand in Eq. (\ref{cons}) vanishes. When the (interacting)
system is slightly biased,  the dot Green functions are not known
exactly. However, the finite bias causes only very small changes
in the Fermi functions $f_{\ell}$ and $f_{r}$,  of the left and of
the right reservoirs,  except  in the range $\mu_{\ell}-\mu_{r}$
around the Fermi energy. Here,
\begin{eqnarray}
f_{\ell}(\omega )=\frac{1}{e^{\beta (\omega -\mu_{\ell})}+1},\ \
f_{r}(\omega )=\frac{1}{e^{\beta (\omega -\mu_{r})}+1}.
\end{eqnarray}
Hence, we expect the integrand in Eq. (\ref{cons}) to be dominated
by contributions from that vicinity of the Fermi energy. If the
integrand in Eq. (\ref{cons}) varies slowly with the frequency
there, then the vanishing of the integral would also imply the
vanishing of the integrand, namely,
\begin{eqnarray}
G^{<}_{dd}=\Sigma^{<}_{\rm
ext}\frac{G^{R}_{dd}-G^{A}_{dd}}{\Sigma^{R}_{\rm
ext}-\Sigma^{A}_{\rm ext}}.\label{GDKELDYSH}
\end{eqnarray}
In some cases, \cite{jauho} this equation follows from the
`wide-band-approximation' which neglects the $\omega$-dependence
of the resonance width $\Im\Sigma_{\rm ext}^{A}$.  Equation
(\ref{GDKELDYSH}) is used  to eliminate the dot Keldysh Green
function from the expressions for the currents. It should be
emphasized (see Appendix A4) that the relationship Eq.
(\ref{GDKELDYSH}) is {\it exact} for the un-biased system. This
point is important for the calculation of the persistent current,
for which one has to keep the contributions of all frequencies. We
note in passing that the sum $I_{2}+I_{4}$ vanishes identically as
checked by an explicit calculation.

The next step taken in Appendix A is to employ the partial
currents in order to obtain the transport current [Eq.
(\ref{Itran})] and the persistent current [Eq. (\ref{Ipc})]. The
former is obtained using the wide-band approximation, in which the
frequency-dependence of the self-energies is suppressed (see
Appendix A3 for details),
\begin{eqnarray}
I&=&e\int\frac{d\omega}{2\pi}(f_{r}-f_{\ell}) \Bigl \{T_{B}\Bigl
(1+G^{R}_{dd}\Sigma^{R}_{\rm
ext}\nonumber\\
&+&G^{A}_{dd}\Sigma^{A}_{\rm ext}+\Sigma^{R}_{\rm
ext}\Sigma^{A}_{\rm
ext}\frac{G^{R}_{dd}-G^{A}_{dd}}{\Sigma^{R}_{\rm
ext}-\Sigma^{A}_{\rm ext}}\Bigr )\nonumber\\
&+&4\Gamma_{\ell}\Gamma_{r}X_{B}\frac{G^{R}_{dd}-G^{A}_{dd}}{\Sigma^{R}_{\rm
ext}-\Sigma^{A}_{\rm ext}}
+\sqrt{T_{B}\Gamma_{\ell}\Gamma_{r}X_{B}}2\cos\Phi\nonumber\\
&&\times\Bigl ( G^{R}_{dd}+G^{A}_{dd}+(\Sigma^{R}_{\rm
ext}+\Sigma^{A}_{\rm
ext})\frac{G^{R}_{dd}-G^{A}_{dd}}{\Sigma^{R}_{\rm
ext}-\Sigma^{A}_{\rm ext}}\Bigr )\Bigr \}.\label{ITRAN}
\end{eqnarray}
%Here, $f_{\ell}$ and $f_{r}$ are the Fermi distributions in the
%left and right reservoirs, respectively,
%\begin{eqnarray}
%f_{\ell}(\omega )=\frac{1}{e^{\beta (\omega -\mu_{\ell})}+1},\ \
%f_{r}(\omega )=\frac{1}{e^{\beta (\omega -\mu_{r})}+1}.
%\end{eqnarray}
The transport current  consists of three parts: The first term in
the curly brackets of Eq. (\ref{ITRAN}) is the current flowing
through the reference arm (the lower arm of the interferometer in
Fig. \ref{fig1}), `dressed' by the processes in which the
electrons travel around the ring, as is manifested by the
appearance of the dot Green functions. Here, $T_{B}$ [see Eq.
(\ref{TB})]  is the transmission coefficient of the reference
branch, when the upper arm of the interferometer is cut off. The
second term in the curly brackets of Eq. (\ref{ITRAN}) is the
current flowing through the interferometer arm containing the
quantum dot; Here $\Gamma_{\ell}$ and $\Gamma_{r}$ [see Eq.
(\ref{GAMMAL})] are the partial linewidths on the dot, caused by
the couplings to the leads, and
\begin{eqnarray}
X_{B}=1-T_{B}\frac{(\gamma_{\ell}+\gamma_{r})^{2}}{4\gamma_{\ell}\gamma_{r}},
\label{XB}
\end{eqnarray}
where $\gamma_{\ell}$ and $\gamma_{r}$ [see Eq. (\ref{gammal})]
are the partial linewidths on the reference site, caused by the
couplings to the leads. (Note that when these are symmetric,
$\gamma_{\ell}=\gamma_{r}$, $X_{B}$ becomes equal to the
reflection coefficient of the reference branch, $R_{B}=1-T_{B}$.)
The last term in Eq. (\ref{ITRAN}) results directly from
interference, since it necessitates transmission through both arms
of the interferometer, as is manifested by the product
$\sqrt{T_{B}\Gamma_{\ell}\Gamma_{r}}$ there.

Several comments on the result (\ref{ITRAN}) are called for: 1.
The transport current $I$ is {\it even} in the Aharonov-Bohm flux
as it should be, obeying the Onsager symmetry. \cite{buttions}
This happens because \cite{avi} the dot Green function
$G_{dd}^{R}$ and $\Sigma^{R}_{\rm ext}$ are even functions of
$\Phi$, due to additive contributions (with equal amplitudes) from
clockwise and counterclockwise motions of the electron around the
ring.  2. When the electronic system is free of interactions, the
`external' self-energy part $\Sigma_{\rm ext}$ constitutes the
entire self-energy of the dot Green function, namely,
\begin{eqnarray}
G^{R0}_{dd}=\frac{1}{\omega -\epsilon_{d}-\Sigma_{\rm ext}^{R}},
\label{GR0}
\end{eqnarray}
where $\epsilon_{d}$ is the energy of the localized level on the
dot, and the superscript `0' denotes the absence of interactions.
In that case $(G^{R0}_{dd}-G^{A0}_{dd})/(\Sigma^{R}_{\rm
ext}-\Sigma^{A}_{\rm ext})=G^{R0}_{dd}G^{A0}_{dd}$, and the
transport current becomes
\begin{eqnarray}
I^{0}=e\int\frac{d\omega}{2\pi}(f_{r}(\omega ) -f_{\ell}(\omega )
)T^{0}(\omega ) \simeq\frac{e^{2}}{2\pi}T^{0}(0) V,\label{I0sof}
\end{eqnarray}
where $V$ is the potential difference on the interferometer, and
$T^{0}$ is the transmission coefficient of the non-interacting
ring,
\begin{eqnarray}
T^{0}(\omega )&=&|G^{R0}_{dd}(\omega
)|^{2}\nonumber\\
&&\times|\sqrt{T_{B}}(\omega
-\epsilon_{d})e^{i\Phi}+2\sqrt{\Gamma_{\ell}\Gamma_{r}X_{B}}|^{2}.\label{t0}
\end{eqnarray}
Equation (\ref{t0})  resembles the {\it two-slit} formula, as it
consists of the absolute value squared of the sum of two terms:
The one related to the transmission amplitude of the reference arm
(having the factor $\sqrt{T_{B}}$) and the other which is related
to the transmission amplitude through the dot (as expressed by
$\sqrt{\Gamma_{\ell}\Gamma_{r}}$), with the Aharonov-Bohm phase
factor multiplying one of those. However, in contrast to the
two-slit formula, here both terms are {\it real}, resulting in an
expression which is even in the flux. This aspect of the
transmission has been discussed in great detail in Refs.
\onlinecite{avi} and \onlinecite{weidenmuller}. In the next
section, we  find that it persists also when the electrons on the
dot are exposed to external radiation. 3. For general values of
the flux $\Phi$, the interaction-free transmission (\ref{t0}) {\it
does not} show the Fano anti-resonances, at which the transmission
vanishes (although the line-shape will be asymmetric). The reason
is that when $\Phi\neq 0$ or $\Phi\neq\pi$, the interference
between the two arms of the interferometer can never be made
completely destructive, as was noted in Ref. \onlinecite{Fano}. On
the other hand, finite values of the flux do not prevent the
transmission from achieving the unitary limit. Inspection of Eq.
(\ref{t0}) in conjunction with the explicit expressions for the
external self-energy, Eqs. (\ref{sigmaRWB}), shows that the
maximal value for the transmission, $T^{0}=1$, is reached when the
interferometer is symmetric, i.e., when $\Gamma_{\ell}=\Gamma_{r}$
and $\gamma_{\ell}=\gamma_{r}$, the local level on the dot becomes
a resonance, i.e., $\omega -\epsilon_{d}-\Re\Sigma_{\rm
ext}^{R}=0$, and the Aharonov-Bohm flux takes the particular value
$\cos\Phi =-T_{B}/(1+R_{B})$.

We next turn to the computation of the circulating current in an
open ring, Eq. (\ref{Ipc}). In the case of non-interacting
electrons, that  current  has been the subject of several studies.
\cite{claudio,buttiker1,avron,mello} Here we generalizes those
calculations to the case where the electrons experience
interactions on the quantum dot.

Inserting the expressions for the partial currents into Eq.
(\ref{Ipc}) (see Appendix A4 for details) we find that the
circulating current consists of two contributions. The first one,
[see Eq. (\ref{IpcA})], is related to the {\it sum} of the two
electronic distributions, $f_{\ell}+f_{r}$. It therefore flows
even when the interferometer is un-biased, and $f_{\ell}=f_{r}$.
The second contribution, [see Eq. (\ref{IaA})], arises only when
the system is biased, being related to the difference
$f_{\ell}-f_{r}$, and only when, in addition, the couplings of the
dot and/or the couplings of the reference site to the
interferometer are not equal, namely, when $i_{\ell}\neq i_{r}$
and/or $j_{\ell}\neq j_{r}$ [see Eqs. (\ref{coupling})]. Both
contributions are induced by the Aharonov-Bohm flux and hence are
proportional to $\sin\Phi$. However, the second term seems to be
not as interesting  as the first. We  therefore omit any further
consideration of that part of the circulating current, and focus
only the first contribution, which reads \cite{prl}
\begin{eqnarray}
I_{pc}&=&e\int\frac{d\omega}{i\pi}\frac{f_{\ell}+f_{r}}{4}\Bigl
[\frac{\partial\Sigma^{R}_{\rm
ext}}{\partial\Phi}G^{R}_{dd}-cc\Bigr ].\label{Ipcsof}
\end{eqnarray}
It is interesting to note that Eq. (\ref{Ipcsof}) averages the
flux-derivative of the external self-energy over energy, with
weights containing the densities of electrons and single-particle
states with that energy (which are contained in $G_{dd}$). The
flux-derivative of $G_{dd}$ does {\em not} appear. This is
reminiscent of the equilibrium case, where the persistent current
is given by the flux-derivative of the energies, weighed by the
electronic populations, without the appearance of the
flux-derivative of those, (see, for example, Ref.
\onlinecite{aronov}). Since both $\Sigma^{R}_{\rm ext}$ and
$G^{R}_{dd}$ are even in $\Phi$,  $I_{pc}$ is odd in $\Phi$, as it
should.

It is sometimes useful to discuss the properties of an open
electronic system in the language of scattering theory,
\cite{buttiker1,mello} employing the concept of `transmission
phases', or the Friedel phase. Such a description is particularly
useful in the case of  interaction-free electrons. Indeed, by
manipulating Eqs. (\ref{GR0}) and (\ref{Ipcsof}), one obtains that
the persistent current of such a system, $I^{0}_{pc}$, is given by
\begin{eqnarray}
I_{pc}^{0}=e\int \frac{d\omega}{\pi} \frac{f_{\ell}(\omega
)+f_{r}(\omega )}{2} \frac{\partial\delta^{0}(\omega
)}{\partial\Phi},\label{Ipc0}
\end{eqnarray}
where $\delta^{0}$ is the phase of the retarded Green function
$G^{R0}=(G^{A0})^\ast$,
\begin{eqnarray}
\tan\delta^{0}(\omega )=-\frac{\Im\Sigma^{R}_{\rm ext}}{\omega
-\epsilon_{d}-\Re\Sigma^{R}_{\rm ext}}.\label{tan}
\end{eqnarray}
Hence, in a steady-state situation, the persistent current of
non-interacting electrons is related to the variation of the
transmission-phase with the Aharonov-Bohm flux. (See Ref.
\onlinecite{avron} for a different derivation of this result.)
This variation replaces the variation of the eigen-energies with
the flux in the equilibrium situation as the origin of the
persistent current. \cite{aronov}

\section{Radiation effects on the transport
current}

The coupling between the electrons residing on the dot and a sonic
source may be described by a linear, local, electron-phonon
interaction, \cite{holstein}
\begin{eqnarray}
{\cal H}_{\rm el-ph}=\sum_{\bf q}\alpha_{\bf q}(b^{\dagger}_{\bf
q}-b_{\bf q})d^{\dagger}d,\label{HEP}
\end{eqnarray}
in which $\alpha_{\bf q}=-\alpha_{-{\bf q}}=-\alpha^{\ast}_{\bf
q}$ is the electron-phonon coupling and $b^{\dagger}_{\bf q}$ is
the creation operator for the boson of wavevector ${\bf q}$. To
study the transport current in the presence of such an
interaction, one has to add ${\cal H}_{\rm el-ph}$ to the
Hamiltonian Eq. (\ref{ham}), together with the free Hamiltonian of
the boson excitations, to compute the dot Green function, and then
to use it in Eq. (\ref{ITRAN}). In the case of a linear
electron-phonon interaction, one is able to obtain an approximate
form for the Green function $G_{dd}$, by assuming that the
external self-energy does not depend on the frequency.
\cite{glazman,wingreen} (For a numerical solution in the presence
of an {\it equilibrium} phonon source, see Ref.
\onlinecite{bonca}.) This is a valid approximation, since the
small potential difference, temperature, etc.,  restrict the
frequency-integration in Eq. (\ref{ITRAN}) to a narrow range
around the common Fermi energy of the two reservoirs.  The
explicit expression for $\Sigma^{R}_{\rm ext}$, valid for the case
of an Aharonov-Bohm interferometer, is given in Eq.
(\ref{sigmaRWB}).

The Green function of the dot, which takes into account the
electron-phonon coupling (\ref{HEP}), was found in Refs.
\onlinecite{glazman}, \onlinecite{wingreen}, and
\onlinecite{jauho}. Here we extend their result to include the
effect of the reference arm and to allow for a finite electronic
occupation, $n_{d}$,  on the dot. The resulting form is then
\begin{eqnarray}
G_{dd}^{R}(\omega )=&-&iK\Bigl
[(1-n_{d})\int_{0}^{\infty}dte^{i(\omega
-\epsilon_{d}-\Sigma^{R}_{\rm
ext})t}e^{\Psi (t)}\nonumber\\
&+&n_{d}\int_{0}^{\infty}dte^{i(\omega
-\epsilon_{d}-\Sigma^{R}_{\rm ext})t}e^{\Psi (-t)}\Bigr ].
\label{Gddpho}
\end{eqnarray}
In the non-equilibrium case, $n_{d}$ is determined by both the
acoustic intensity and the relaxation processes. The on-site
energy on the dot, $\epsilon_{d}$, is now renormalized by the
polaron shift, $\epsilon_{P}=\sum_{\bf q}|\alpha_{\bf
q}|^{2}/\omega_{q}$, where $\omega_{q}$ denotes the phonon
frequency. Since this renormalization is temperature- and
flux-independent, it will be omitted. The other phonon variables
are contained in $K$, the Debye-Waller factor, and in $\Psi (t)$.
Explicitly,
\begin{eqnarray}
K&=&\exp[-\sum_{\bf q}\frac{|\alpha_{\bf
q}|^{2}}{\omega_{q}^{2}}(1+2N_{q})],\nonumber\\
\Psi (t)&=&\sum_{\bf q}\frac{|\alpha_{\bf q}|^{2}}{\omega_{q}^{2}}
[N_{q}e^{i\omega_{q}t}+(1+N_{q})e^{-i\omega_{q}t}],
\end{eqnarray}
where  $N_{q}=\langle b^{\dagger}_{\bf q}b_{\bf q}\rangle$ is the
phonon occupation of the ${\bf q}$-mode, which is not necessarily
the thermal equilibrium one, but may be tuned externally.

Perhaps the simplest way to access the effect of the acoustic
coupling  is by expanding $G_{dd}$ in the electron-phonon coupling
$|\alpha_{\bf q}|^{2}$,
\begin{eqnarray}
G^{R}_{dd}(\omega )&=&KG^{R0}_{dd}(\omega )+\sum_{s=\pm }\sum_{\bf
q}C^{s}_{\bf q}G^{R0}_{dd}(\omega +s\omega_{q}), \label{expansion}
\end{eqnarray}
where the interaction-free Green function $G^{R0}_{dd}$ is given
in Eq. (\ref{GR0}). Here, $s=\pm$, and
\begin{eqnarray}
C^{+}_{\bf q}=\frac{|\alpha_{\bf
q}|^{2}}{\omega_{q}^{2}}(N_{q}+n_{d}),\  C^{-}_{\bf
q}=\frac{|\alpha_{\bf q}|^{2}}{\omega_{q}^{2}}(1+N_{q}-n_{d}).
\end{eqnarray}
For a weak electron-phonon coupling, the Debye-Waller factor is
\begin{eqnarray}
K\simeq 1-\sum_{s=\pm }\sum_{\bf q}C^{s}_{\bf q}.\label{expK}
\end{eqnarray}
However, it is instructive to keep the Debye-Waller factor $K$,
which multiplies the zero-order term in the expansion
(\ref{expansion}) (and, in principle, all other terms in the
expansion) in its implicit form, in order to demonstrate its role
in diminishing {\it all} contributions to the current, and not
only those arising from interference. \cite{I02}

It is thus seen that the dot Green function in the presence of the
electron-phonon coupling may be written as a series of terms in
which there appear the interaction-free Green functions, with
their frequency argument shifted by the phonon frequencies,
\cite{jonson,argentina} each multiplied by the relevant phonon
occupation numbers. Hence, it is quite obvious that the transport
current will have a similar form. Indeed, upon inserting the
result (\ref{expansion}) into the expression for the transport
current, Eq. (\ref{ITRAN}), one finds
\begin{eqnarray}
I&=&e\int\frac{d\omega}{2\pi}(f_{r}(\omega )-f_{\ell}(\omega
))T^{\rm rad}(\omega ),
\end{eqnarray}
in which the transmission of the irradiated interferometer,
$T^{\rm rad}$, is
\begin{eqnarray}
T^{\rm rad}(\omega )= KT^{0}(\omega )+\sum_{s=\pm}\sum_{\bf
q}C^{s}_{\bf q}T^{0}(\omega +s\omega_{q}),\label{trad}
\end{eqnarray}
and the interaction-free transmission is given in Eq. (\ref{t0}).
It is seen that the processes contained in $T^{0}(\omega
+s\omega_{q})$, compensate partially for the detrimental effect of
the Debye-Waller factor, $K$. We will encounter a similar
situation in the discussion of the circulating current. Since we
are operating in the linear response regime, it suffices to study
the result (\ref{trad}) at the Fermi energy, namely, at zero
frequency in our notations.

Let us first consider the radiation effect on the transport
through the ring in the unitary limit, namely when $T^{0}(0)=1$.
This situation, as  mentioned above, occurs for a symmetric ring,
when $\epsilon_{d}+\Re\Sigma^{R}_{\rm ext}=0$ and  $\cos\Phi
=-T_{B}/(1+R_{B})$. Under these conditions
\begin{eqnarray}
T^{0}(s\omega_{q})\Big |_{\rm res}
=1-R_{B}\frac{\omega_{q}^{2}}{(\Im\Sigma_{\rm
ext}^{R})^{2}+\omega_{q}^{2}}.
\end{eqnarray}
Inserting this into Eq. (\ref{trad}), and using Eq. (\ref{expK}),
yields
\begin{eqnarray}
&&T^{\rm rad}(0 )\Big |_{\rm res}=1-R_{B}\sum_{s=\pm}\sum_{\bf
q}C^{s}_{\bf q}\frac{\omega_{q}^{2}}{(\Im\Sigma_{\rm
ext}^{R})^{2}+\omega_{q}^{2}}\nonumber\\
&=&1-R_{B}\sum_{\bf q}\frac{|\alpha_{\bf
q}|^{2}}{\omega_{q}^{2}}(1+2N_{q})\frac{\omega_{q}^{2}}{(\Im\Sigma_{\rm
ext}^{R})^{2}+\omega_{q}^{2}}.\label{traduni}
\end{eqnarray}
At resonance, the transmission is independent of the electronic
occupation on the dot. The coupling with the bosons reduces the
transmission at resonance, the more so as the intensity of the
boson source in a certain frequency range increases; It is
interesting to note, however, that this effect becomes smaller as
the reflection coefficient of the reference arm decreases (and
therefore the current tends to go mainly through that arm).

To study the effect of the radiation in the general case, it is
convenient to present the transmission $T^{\rm rad}$ in the form
\begin{eqnarray}
T^{\rm rad}(0)&=&T^{0}(0)+\frac{1}{2}\sum_{\bf q}A^{-}_{\bf
q}\Bigl
[T^{0}(\omega_{q})-T^{0}(-\omega_{q})\Bigr ]\nonumber\\
&+ &\frac{1}{2}\sum_{\bf q}A^{+}_{\bf q}\Bigl
[T^{0}(\omega_{q})+T^{0}(-\omega_{q})-2T^{0}(0)\Bigr
],\label{IRAD}
\end{eqnarray}
where $-2T^{0}(0)$ comes from the Debye-Waller factor. Here
$A_{\bf q}^{+}$ is directly proportional to the radiation
intensity, while $A_{\bf q}^{-}$ does not depend on it.
Explicitly,
\begin{eqnarray}
A^{+}_{\bf q}&=&C^{+}_{\bf q}+C^{-}_{\bf q}=\frac{|\alpha_{\bf q}|^{2}}
{\omega_{q}^{2}}(1+2N_{q}),\nonumber\\
A^{-}_{\bf q}&=&C^{+}_{\bf q}-C^{-}_{\bf q}=\frac{|\alpha_{\bf
q}|^{2}}{\omega_{q}^{2}}(2n_{d}-1).\label{APM}
\end{eqnarray}
Equation (\ref{IRAD}) shows that by shining a beam of bosons at a
certain frequency range,  the transport current increases linearly
with  the intensity of the beam, as long as the latter is not too
large. For example, when the interferometer is far from resonance,
namely when $|\epsilon_{d}|\gg\Gamma_{0}$, where
$\Gamma_{0}=\Gamma_{\ell}+\Gamma_{r}$, ($\Gamma_{0}$  is the width
of the resonance level of the quantum dot itself, when it is
disconnected from the reference arm), we find that the
transmission, to lowest order in $\Gamma_{0}/|\epsilon_{d}|$
becomes
\begin{eqnarray}
&&T^{\rm rad}(0)\Big |_{\rm off\ res}=T_{B}
-\sqrt{T_{B}R_{B}}\Bigl (T_{B}+(1+R_{B})\cos\Phi \Bigr )\nonumber\\
&&\times\Bigl [\frac{\Gamma_{0}}{\epsilon_{d}}\Bigl (1+\sum_{\bf
q} A_{\bf
q}^{+}\frac{\omega_{q}^{2}}{\epsilon_{d}^{2}-\omega_{q}^{2}}\Bigr
)+\Gamma_{0}\sum_{\bf q} A_{\bf
q}^{-}\frac{\omega_{q}}{\epsilon_{d}^{2}-\omega_{q}^{2}}\Bigr ].
\end{eqnarray}
Of particular interest is the point that the magnitude of the
interference term can be controlled by coupling the dot to a sonic
source. The other factor, $A_{\bf q}^{-}$, may change sign
depending on the relative location of the on-site energy on the
dot and the Fermi level, but its magnitude cannot vary much,
$-1\leq 2n_{d}-1\leq 1$.

\section{Radiation effects  on the circulating
current}

The subtle effect that electron-phonon interactions may have on
interference-related properties of electrons has been invoked a
long time ago by Holstein, \cite{holstein1} in his theory of the
Hall effect in hopping conduction.  Holstein proposed that in
order to capture the Hall effect, it is necessary to consider
processes where the amplitude of the direct electron tunneling
between two `sites' around which the electronic wave functions are
localized, interferes with an indirect tunneling amplitude,
through an intermediate third site. Moreover, that interference
must involve energy-conserving electron transitions to and from
the intermediate site, which are assisted by phonons. It turns out
that this `Holstein process'  has intriguing consequences for the
persistent current in {\it electronically-isolated}
interferometers. \cite{aronov} Since it is of interest to compare
the radiation effect on persistent currents in isolated and in
open rings, we  begin   this section with a  brief summary of the
Holstein process and its consequences for the isolated system, and
then  analyze the situation in an open ring.

The Holstein mechanism  can be explained in a somewhat technical
language as follows. Under hopping conduction conditions,
transport can be related to {\it transition probabilities}.
Imagine now the transition probability per unit time, $P_{ij}$, to
tunnel from the electronic state localized at $i$ to that
localized at $j$. When the system is subject to a constant
magnetic field, the tunneling amplitude between $i$ and $j$ is
multiplied by the magnetic phase acquired from the field along the
path $i-j$. Upon taking the absolute value squared of such an
amplitude to obtain $P_{ij}$ due to direct hopping alone, the
result is independent of the magnetic field. Now let us add to the
direct tunneling amplitude between $i$ and $j$ the amplitude for
indirect tunneling, for example, the  path $i-\ell -j$, where
$\ell$ denotes an intermediate site. The transition probability
now depends on the total, gauge-invariant, magnetic flux enclosed
by the two paths (i.e., the Aharonov-Bohm phase). However, it is
an {\it even} function of the magnetic phase, as the tunneling
amplitudes themselves can always be  chosen to be real. As such,
this transition probability cannot lead to a dc Hall conduction,
which is {\it odd} in the field. This line of argument shows that,
technically speaking, an imaginary contribution to at least one of
the transition amplitudes  is required in order to render a term
odd in the magnetic phase in the transition probability.

Where can this imaginary part come from? Holstein \cite{holstein1}
argued that when electron-phonon processes are taken into account,
the intermediate state becomes in fact a continuum of energy
states, consisting of the intermediate electronic energy, and the
continuum of phonon energies. This continuum suffices to supply
the required imaginary contribution. Roughly speaking, when
electron-phonon interactions are accounted for, the tunneling
amplitude for the indirect path acquires,  for $\epsilon_{\ell} >
\epsilon_{i}$, terms such as
\begin{eqnarray}
J_{i-\ell-j}&&\sim\sum_{\stackrel{n_{\bf q},{\bf q}}{n_{{\bf
q}'},{\bf q}'}} \frac{\langle \ell, n_{{\bf q}} - 1, n_{{\bf q'}}
|V| j, n_{{\bf q}}, n_{{\bf q}'} \pm 1 \rangle}
{\epsilon_{i}-\epsilon_{j}\mp\omega_{q'}+i\eta}\nonumber\\
&&\times \langle j, n_{{\bf q}}, n_{{\bf q}'} \pm 1|V|i, n_{{\bf
q}}, n_{{\bf q}'}\rangle .\label{comb}
\end{eqnarray}
%
%$$J_{i-\ell-j}~~~\sim  \sum_{n_{{\bf q}},{\bf q},n_{{\bf q}'},{\bf
%q}'}~~~~~~~~~~~~~~~~~~~~~~~~~~~~~~~~~~~~~~~~~~$$
%\begin{equation}
%\frac { \langle \ell, n_{{\bf q}} - 1, n_{{\bf q'}} |V| j, n_{{\bf
%q}}, n_{{\bf q}'} \pm 1 \rangle \langle j, n_{{\bf q}}, n_{{\bf
%q}'} \pm 1|V|i, n_{{\bf q}}, n_{{\bf
%q}'}\rangle}{\epsilon_{i}-\epsilon_{j} \mp \omega_{q'} + i \eta}.
%\label{comb}
%\end{equation}
Here, $\epsilon_{i}$, etc.,  denotes  electronic site energies,
$\eta\rightarrow 0^{+}$, $\omega_{q}$ and $\omega_{q'}$ are boson
energies, and $n_{\bf q}$ and $n_{{\bf q}'}$ are  the quantum
numbers of the ${\bf q}-$ and ${\bf q}'-$ mode occupations. In Eq.
(\ref{comb}), $V$ is the operator that transfers the electron
between sites, and at the same time may cause the phonon states to
change, obtained after an appropriate \cite{holstein1} unitary
transformation on the electron-boson Hamiltonian, Eq. (\ref{HEP}).
Since the intermediate state now lies in a {\it continuum} of
energies, the infinitesimal part $\eta$~ leads to a a {\it finite}
imaginary contribution, provided that the sum of energies in the
denominator vanishes, namely, when there is an exact energy
conservation, as would be needed to make a {\it real} transition
\cite{miller} between the initial and intermediate states of the
process. We emphasize, however, that the boson created/destroyed
in going from $i$ to $j$ is only virtual, exactly {\it the same}
boson is destroyed/created in going from $j$ to $\ell$. This exact
identity is necessary for {\em incoherent} phonons in order to
retain phase coherence \cite{ady} with the direct process from $i$
to $\ell$. More technically,  one uses the relation $ 1/(x+i\eta
)={\cal P}/x-i\pi \delta (x),$ where ${\cal P}$ denotes the
principal part. The delta-function term within the infinite sum
over the phonon modes gives rise to the required finite imaginary
contribution. The resulting imaginary part in $J_{i-\ell -j}$
yields  a term odd in the flux in the transition probability. It
is worth noting that the energy-conserving process occurs here in
the intermediate state of the perturbation theory [of which Eq.
(\ref{comb}) is the lowest term] for the combined amplitudes.
Recently, this unique process has been proposed as the origin of
the anomalous Hall effect in ferromagnetic semiconductors.
\cite{balents}

The argument above exemplifies the necessity for one resonant
process. However, in fact the Holstein process requires at least
two resonant electron-phonon processes. This can be explained  as
follows: The three electronic energies involved in the indirect
tunneling and their differences are in general all different.
Hence, at least one phonon (the one denoted above by $q$) is
needed to supply the energy difference
$\epsilon_{i}-\epsilon_{\ell}$ between the initial and final
electronic states. The second phonon ($q'$ above) appears in the
intermediate process, as explained above. We will come back to
this point in the following. The phonon-assisted indirect
amplitude, Eq. (\ref{comb}), gives rise also to a contribution
which is {\it even} in the field (coming from the principal part).
That contribution does not require exact energy-conservation
within the intermediate state of the perturbation energy (it does
however, require the phonon supplying the energy difference
between the initial and final electronic states).

The fact that the transition probability per unit time for an
electron to hop between two sites may include a term which is odd
in the Aharonov-Bohm flux (in addition to the term even in the
flux) has an immediate result: detailed balance is broken  even at
thermal equilibrium. Stated in terms of transition probabilities,
$P_{ij}- P_{ji}\neq 0$, and the difference is {\it odd} in the
magnetic flux. To appreciate the outcome of this observation, let
us focus our attention on a triad of three sites, $i$, $j$, and
$\ell$, the smallest cluster in which the doubly-resonant
transitions can take place. The transition probability to go from
site $i$ to site $j$, $P_{ij}$, (which includes also the indirect
processes via site $\ell$), and the transition probability to go
from that site to site $\ell$ (now also through the intermediate
site $j$), $P_{i\ell}$, are such that
\begin{eqnarray}
P_{ij}+P_{i\ell}=P_{ji}+P_{\ell i},
\end{eqnarray}
so that charge balance is maintained at the electronic site $i$.
However, since $P_{ij}\neq P_{ji}$, there is a net current
circulating around the triad,  proportional to $P_{ij}-P_{ji}$,
and therefore arising from the Holstein process. That current is
additional to the persistent current flowing in this system in the
absence of the coupling to the phonon source. In fact, it has been
found \cite{aronov} that it is always flowing in the reverse
direction! (The direction of the current in the triad is
determined by delicate effects related to the location of the
Fermi level with respect to the site energies, etc.) This current
has been therefore termed `counter-current'. When the full
transition probabilities, including the terms even and odd  in the
magnetic flux,  are  used in the proper rate equations to find the
current, the resulting conductivity tensor satisfies the Onsager
relations. \cite{oldPRL}

Having related the doubly-resonant processes of Holstein to the
persistent current, it is worthwhile to re-examine the resonance
conditions from the point of view of coherence. As we have pointed
above, and as is borne out by the full calculation, \cite{aronov}
one of the two phonons is common to both  interfering tunnelling
paths, thus retaining their coherence, \cite{ady} while the other
is,  as explained above, absorbed and re-emitted by one of the
paths, again retaining coherence with the other path. Hence,
albeit the fact that the Holstein mechanism also involves a real,
energy-conserving, electron-phonon transition, it still
contributes in a non-trivial way to the persistent current.
However, since this contribution arises from `real' processes, it
requires `real' phonon modes, namely, non-zero temperatures. One
therefore expects that the counter-current will {\it increase}
with the temperature. On the other hand, the counter-current is
also multiplied by the overall damping Debye-Waller factor. Hence,
the resulting temperature dependence of the counter-current is
non-monotonic. \cite{aronov}

When the interferometer is connected by leads to external
electronic reservoirs, the energy levels on the ring acquire
finite widths, given by the imaginary part of the external
self-energies. Then, the effect of the coupling to the sonic
source is modified. While for discrete states, it required  exact
energy conservation  (up to the width introduced by the coupling
to the phonons), here it operates in a finite energy band.
Nonetheless, the radiation introduces again a unique effect, which
goes beyond that of the Debye-Waller exponent. In the present
situation the sonic effect is of a lower order in the
electron-phonon coupling, and may exist even in the $T\rightarrow
0$ limit as will be discussed later.   (For a concise summary of
this result see Ref. \onlinecite{prl}.)

Indeed, inserting the expansion of the dot Green function at small
electron-phonon coupling, Eq. (\ref{expansion}), into our general
result for the circulating current, Eq. (\ref{Ipc}), yields
\begin{eqnarray}
I_{pc}=I_{pc}^{0}+ \Delta I_{pc},\label{Ipcrad}
\end{eqnarray}
where $I_{pc}^{0}$ is the persistent current of the
non-interacting interferometer, given by Eq. (\ref{Ipc0}) above,
and $\Delta I_{pc}$ is the acousto-persistent current, given,
within our approximation,  by
\begin{eqnarray}
&&\Delta I_{pc}=\int \frac{d\omega}{\pi} \frac{f_{\ell}(\omega
)+f_{r}(\omega )}{4}\nonumber\\
&&\times \sum_{\bf q} \Bigl [A^{-}_{q}\frac{\partial
}{\partial\Phi}\Bigl (\delta^{0} (\omega
+\omega_{q})-\delta^{0}(\omega -\omega_{q})\Bigr )\nonumber\\
&+&A^{+}_{q}\frac{\partial }{\partial\Phi}\Bigl (\delta^{0}(\omega
+\omega_{q})+\delta^{0}(\omega -\omega_{q}) -2\delta^{0}(\omega
)\Bigr )\Bigr
 ], \label{ICPH}
\end{eqnarray}
where $\delta^{0}$, the Friedel phase of the non interacting
system, is given in Eq. (\ref{tan}), and $A^{\pm}_{\bf q}$ is
defined  in Eqs. (\ref{APM}). The acousto-induced persistent
current, $\Delta I_{pc}$, consists of two parts: The first term
depends only on the dot's occupation, $n_d$, and its sign may
change according to the relative location of $\epsilon_d$ with
respect to the Fermi energy. The second term in Eq. (\ref{ICPH})
is dominated by the phonon occupations [see Eq. (\ref{APM})], via
$A_{\bf q}^{+}$. [Note that the term $-2\delta^{0}(\omega )$ there
comes from the expansion of the Debye-Waller exponent.]
%, and can be
%therefore included in $I_{pc}^{0}$
Examining this contribution shows that by shining a beam of
phonons of a specific frequency, the magnitude of that term can be
enhanced and controlled experimentally, as long as the temperature
of the electronic system and the intensity of the phonon source
$N_q$ are low enough to retain coherent motion of the electrons.
(The intensity is also limited in the present calculation by the
assumption of weak electron-phonon  coupling; However, there is no
conceptual difficulty to extend the calculation to stronger
values.) Similar considerations apply to photons. Both the precise
magnitude of these effects and the above bounds depend on the
detailed geometry of the dot and on the acoustic (or
electromagnetic) mismatch.

It is important to appreciate the difference between this result
and the corresponding one found for the isolated ring. In the
isolated ring, the Holstein process \cite{holstein} required the
emission (absorption) of a specific phonon, with the exact
excitation energy of the electron on the ring. In the present
case, the coupling to the leads turns the bound state into a
resonance, with a width $\Gamma_0$ which vanishes when the ring is
decoupled from the leads. As a result, there is always some
overlap between the tail of the Green function
$G_{dd}^{R0}(\omega)$ and the Fermi distribution $f(\omega)$,
yielding contributions from Holstein-like processes via phonons
with many (including very low) energies. Indeed, each contribution
to $\Delta I_{pc}$ contains the phase $\delta^0(\omega)$, which
vanishes with $\Gamma_0$ ($\delta^0 \sim \Gamma_0/|\epsilon_d|$
far from the resonance). In particular, this results in a non-zero
$\Delta I_{pc}$ even at zero temperature: In that limit, if
$\epsilon_d<\mu_\ell=\mu_r=0$, then $n_d=1$. Even with no phonons,
$N_q=0$, the square brackets in Eq. (\ref{ICPH}) become
proportional to $\partial [
\delta^0(\omega+\omega_q)-\delta^0(\omega)]/\partial\Phi$,
reflecting processes which begin by an emission of phonons. None
of this remains for the isolated ring, when $\Gamma_0=0$.
% It
%should however be noted that the emission and re-absorption (or
%{\it vice versa}) are with {\it the same} phonon. As discussed
%before, the existence of the coupling to the leads allows the
%Holstein process to apply in second-order in the electron-phonon
%coupling. In the closed ring, having discrete electronic levels,
%another phonon was needed for the (real) hopping process yielding
%the term odd in the flux.

To obtain explicit expressions, we now evaluate the frequency
integration appearing in Eq. (\ref{ICPH}). Since we are operating
within the linear response regime, the voltage is not essential to
our effect and we may safely write in Eq. (\ref{ICPH})
$f_{\ell}(\omega )=f_{r}(\omega )\equiv f(\omega)$ . Furthermore,
we take the electronic temperature to be low compared to all other
energies, so that $f(\omega) \approx \Theta(-\omega)$. We also
take the typical phonon frequency to be  much smaller than the
large band-width in the leads. With these approximations the
frequency-integration in Eq. (\ref{ICPH}) is easily performed, to
yield
\begin{eqnarray}
\Delta I_{pc}&=&\frac{\Gamma_{0}}{4\pi}\sin\Phi\sum_{\bf q}\Bigl
[A^{+}_{\bf q}\Bigl (F(\omega_{q})+F(-\omega_{q})-2F(0)\Bigr
)\nonumber\\
&+&A^{-}_{\bf q}\Bigl (F(\omega_{q})-F(-\omega_{q})\Bigr )\Bigr ],
\end{eqnarray}
where $F(\omega )$ is given in terms of $\delta^{0}(\omega )$, Eq.
(\ref{tan}),
\begin{eqnarray}
F(\omega )=-\sqrt{T_{B}R_{B}}\delta^{0}(\omega )-T_{B}\ln
|\sin\delta^{0}(\omega )|.
\end{eqnarray}
Note again that the dependence of the acousto-persistent current
on the phonon frequency is determined by the Friedel phase of the
dot at that frequency.  To leading order in the strength of the
electron-phonon coupling, the magnitude of the first term in
$\Delta I_{pc}$ is proportional to $A^{+}_{\bf q}$, and thus grows
linearly with the occupation number of the acoustic modes acting
on the dot, $N_q$. In fact, the acousto-persistent current
contains two types of contributions: the part associated with
$F(0)$, which simply represents the `trivial' Debye-Waller
renormalization of the current, and the novel frequency-dependent
parts, which reflect the change in the persistent current due to
Holstein-like processes.

The above discussion holds at thermal equilibrium. There, the
electron-phonon interaction does not enhance the persistent
current. On the other hand, taking the phonon source out of
equilibrium at a certain frequency range may lead to enhancement.
On a speculative level, remembering that the Debye-Waller exponent
depends on the sum of all phonon occupations (weighted by their
couplings to the electronic levels), while the counter-current
depends only on the phonon occupations of the two resonating
frequencies, one may visualize the following set-up: Suppose that
one shines on the electronic system a high intensity beam of
non-equilibrium phonons (or photons) with a narrow frequency range
around, say, $\omega_{0}$. The counter-current, resulting from
resonant transitions, will be significantly affected by the
non-equilibrium phonons only when $\omega_{0}$ will be close to
the differences $|\epsilon_{i}-\epsilon_{j}|$ or
$|\epsilon_{i}-\epsilon_{\ell}|$. The effect on the Debye-Waller
factor, on the other hand, will be small for a narrow-band beam.
In this way, the counter-current will initially increase with the
intensity of this radiation, until the
Debye-Waller/decoherence/heating effects will take over and the
entire persistent current will disappear.

We have not emphasized in this paper the contribution of
non-Holstein processes [i.e., those arising from the principal
part in Eq. (\ref{comb})]. Such processes are not specific to a
definite phonon frequency, and therefore can not be increased
without heating/decohering the system.

\section{Summary}

We have considered the effect of coupling the electrons to a boson
source on their interference pattern in an Aharonov-Bohm
interferometer, and in particular focused our attention on the
modifications in the transport current and in the circulating
current. In both cases, there appears the overall Debye-Waller
exponent, which reduces the interference term (as well as the
`classical' term), and hence the currents, as the temperature is
raised. All boson modes contribute to the Debye-Waller factor.
This outcome of the coupling to the boson source is not
surprising. However, in both cases, there is an additional
contribution, which is confined to a bounded range of phonon
energies, dictated by the electronic energies.

In the case of hopping conduction, which involves transitions
between discrete localized electronic states that in general
differ in energy, a phonon (common to the two paths) is necessary
to conserve energy in the overall hopping process. \cite{miller}
In the case of an open interferometer, that  phonon is not
necessary, since the electronic states on the two leads form
continua and overlap in energy. To obtain a term odd in the
magnetic field in the hopping regime, another, `second', phonon is
needed, which has {\it to conserve the total energy between the
initial and intermediate states}. \cite{holstein1} The reverse
phonon process (namely, restoring the phonon system back to its
original state) then occurs between the intermediate and final
states, thus retaining phase memory in the overall process (which
can then interfere with another phonon-less path). The
conservation of energy in the intermediate state is a rather
unusual feature, which introduces an imaginary part to the hopping
amplitude  for that path, and hence a nontrivial phase. That phase
was crucial for the theory of the Hall effect in the hopping
regime.
%, much less so in our case.
Here, the process  appears at a lower order in the electron-phonon
interaction, as compared to the situation in isolated rings with
localized electronic states. \cite{aronov} In addition, the
intermediate electronic state acquires a width via coupling to the
leads. Therefore the process may exist even at zero temperature.
This is due to the finite overlap of the intermediate electronic
state with the ``band".

Because this novel contribution to the  currents comes from a
confined range of  boson frequencies, it is expected that by
modulating the intensity of the radiation in that frequency range,
it will be possible to manipulate the magnitude of the currents.
This will require boson intensities low enough to retain the
coherent motion of the electrons. However, the fact that this
unique effect is confined to a rather narrow region of boson
frequencies (while the detrimental Debye-Waller factor comprises
all boson frequencies) gives some hope that such an
acousto-magnetic effect is feasible in experiments.

\begin{acknowledgments}
This project was carried out in a center of excellence supported
by the Israel Science Foundation. It was also partially supported
by the German-Israeli Foundation (GIF), by the U.S. Department of
energy, Office of Science,  through contact No. W-31-109-ENG-38,
and by the German Federal Ministry of Education and Research
(BMBF), within the framework of the German Israeli Project
Cooperation (DIP).
\end{acknowledgments}

\appendix
\section{details of the current calculation}

As is clearly explained in Ref. \onlinecite{langreth} (see also
Ref. \onlinecite{jauho}), the Green functions required in the
Keldysh technique can be found by considering the time-ordered
Green functions, $G^{T}$. The latter satisfy the
frequency-dependent Dyson equation for $G^{T}(\omega )$,
\begin{eqnarray}
G^{T}=G^{0T}+G^{0T}\Sigma^{T}G^{T}.\label{Dyson}
\end{eqnarray}
Then, the retarded ($G^{R}$) and the advanced ($G^{A}$) Green
functions are obtained by replacing $T$ above by $R$ or $A$, while
$G^{<}$ is found according to the rule \cite{langreth}
\begin{eqnarray}
\Bigl (\Sigma G\Bigr
)^{<}=\Sigma^{R}G^{<}+\Sigma^{<}G^{A},\label{rule}
\end{eqnarray}
and similarly for any other product. In the following, we  omit
for brevity the notation $T$ from the time-ordered Green
functions.

\subsection{The calculation of the partial currents}

We now apply the Keldysh method to calculate  the partial
currents, defined in Eq. (\ref{currents}).  Our aim is to express
these currents in terms of the dot Green function, $G_{dd}$.

We begin with the calculation of $I_{1}$. The required Green
functions $G^{<}_{kd}$ and $G^{<}_{dk}$ are obtained as follows.
The equations-of-motion for the temporal Fourier transforms of the
time-ordered counterparts read
\begin{eqnarray}
(\omega -\epsilon_{k})G_{kd}&=&V_{k}G_{dd}+v_{k}G_{0d},\nonumber\\
G_{dk}(\omega
-\epsilon_{k})&=&V_{k}^{\ast}G_{dd}+v_{k}^{\ast}G_{d0}.
\end{eqnarray}
In order to use the rule (\ref{rule}), we re-write these two
equations in the form
\begin{eqnarray}
G_{kd}&=&V_{k}g_{k}G_{dd}+v_{k}g_{k}G_{0d},\nonumber\\
G_{dk}&=&V^{\ast}_{k}G_{dd}g_{k}+v^{\ast}_{k}G_{d0}g_{k},\label{Gdk}
\end{eqnarray}
in which $g_{k}$ is the free Green function of the left lead,
namely
\begin{eqnarray}
g^{R,A}_{k}&=&\frac{1}{\omega \pm i\eta -\epsilon_{k}},\ \
g^{<}_{k}=f_{\ell}(\omega )\Bigl (g^{A}_{k}-g^{R}_{k}\Bigr ).
\label{gk}
\end{eqnarray}
Here $\eta\rightarrow 0^{+}$, and
\begin{eqnarray}
f_{\ell}=\frac{1}{e^{\beta (\omega -\mu_{\ell})}+1},
\end{eqnarray}
is the electron distribution in the left electronic reservoir, and
Eq. (\ref{KELDYSH}) has been employed to obtain $g^{<}_{k}$. Since
we assume that the two leads in Fig. \ref{fig1} are identical
except for being connected to reservoirs of different chemical
potentials, the free Green functions of the right lead are given
by Eqs. (\ref{gk}), with $f_{\ell}$  replaced by $f_{r}$. For
brevity, the dependence on the frequency $\omega$ will be
suppressed in most of the equations.

Using now the rule (\ref{rule}), we find
\begin{eqnarray}
G_{kd}^{<}&=&V_{k}\Bigl
(g^{R}_{k}G^{<}_{dd}+g^{<}_{k}G_{dd}^{A}\Bigr ) +v_{k}\Bigl
(g^{R}_{k}G^{<}_{0d}+g^{<}_{k}G^{A}_{0d}\Bigr ),
\nonumber\\
G_{dk}^{<}&=&V^{\ast}_{k}\Bigl
(G^{<}_{dd}g^{A}_{k}+G_{dd}^{R}g^{<}_{k}\Bigr ) +v^{\ast}_{k}\Bigl
(G^{<}_{d0}g^{A}_{k}+G^{R}_{d0}g^{<}_{k}\Bigr ).\label{GdkK}
\nonumber\\
\end{eqnarray}
Inserting Eqs. (\ref{GdkK}) into $I_{1}$, Eq. (\ref{currents}),
and writing explicitly the couplings $V_{k}$ and $v_{k}$ from Eqs.
(\ref{coupling}), one finds
\begin{eqnarray}
I_{1}&=&e\int\frac{d\omega}{2\pi}\frac{2}{N}\sum_{k}\sin^{2}
k\Bigl
[j_{\ell}^{2}\Bigl (G^{<}_{dd}(g^{R}_{k}-g^{A}_{k})\nonumber\\
&+&g_{k}^{<}(G^{A}_{dd}-G^{R}_{dd})\Bigr
)+j_{\ell}i_{\ell}e^{i\phi_{\ell}}(g^{R}_{k}G^{<}_{0d}+g^{<}_{k}G^{A}_{0d})\nonumber\\
&-&j_{\ell}i_{\ell}e^{-i\phi_{\ell}}(g^{<}_{k}G^{R}_{d0}+g^{A}_{k}G^{<}_{d0})\Bigr
].\label{I10}
\end{eqnarray}
When the explicit expressions for $g_{k}^{A}$, $g_{k}^{ R}$, and
$g_{k}^{<}$ [see Eqs. (\ref{gk})] are inserted into Eq.
(\ref{I10}), it turns out that it  is  useful to define
\begin{eqnarray}
\alpha^{R,A}&=&\frac{2}{N}\sum_{k}g_{k}^{R,A}\sin^{2}k ,
\end{eqnarray}
and
\begin{eqnarray}
\Delta =\alpha^{A}-\alpha^{R}\equiv \frac{4\pi i}{N}\sum_{k}\delta
(\omega -\epsilon_{k})\sin^{2}k.\label{delta}
\end{eqnarray}
With these notations, the partial current $I_{1}$ becomes
\begin{eqnarray}
I_{1}&=&e\int \frac{d\omega}{2\pi}\Bigl (-\Delta j^{2}_{\ell}\bigl
[G_{dd}^{<}+f_{\ell}(G_{dd}^{R}-G_{dd}^{A})\bigr ]\nonumber\\
&+&j_{\ell}i_{\ell}e^{i\phi_{\ell}} \bigl [f_{\ell}\Delta
G^{A}_{0d} +\alpha^{R}G^{<}_{0d}\bigr ]\nonumber\\
&-&j_{\ell}i_{\ell}e^{-i\phi_{\ell}}\bigl
[\alpha^{A}G^{<}_{d0}+f_{\ell }\Delta G^{R}_{d0}\bigr ]\Bigr
).\label{cur1}
\end{eqnarray}

The next step is the find the Green functions $G_{0d}$ and
$G_{d0}$ in terms of the dot Green function $G_{dd}$. This is
accomplished as follows. The equation-of-motion for the
time-ordered Green function $G_{0d}$ reads
\begin{eqnarray}
G_{0d}=g_{0}\Bigl (\sum_{k}v_{k}^{\ast}G_{kd}+\{k\rightarrow
p\}\Bigr ),\label{G0d}
\end{eqnarray}
in which the notations $\{k\rightarrow p\}$ stand for the
analogous sum on the right lead, and $g_{0}$ is the free Green
function on the reference site, with
\begin{eqnarray}
g_{0}^{R,A}=\frac{1}{\omega \pm i\eta -\epsilon_{0}}.
%g_{0}^{<}=0.
\end{eqnarray}
Since the bare reference site is not coupled to any electronic
reservoir, the free Keldysh Green function for that site vanishes,
\begin{eqnarray}
g^{<}_{0}=0.
\end{eqnarray}
Making use of Eqs. (\ref{Gdk}) we have
\begin{eqnarray}
G_{0d}^{A}&=&g^{A}_{0}\Bigl [\sum_{k}v_{k}^{\ast}g_{k}^{A}\Bigl
(V_{k}G_{dd}^{A}+v_{k}G_{0d}^{A}\Bigr )+\{k\rightarrow p\}\Bigr
],\nonumber\\
&&
\end{eqnarray}
which yields
\begin{eqnarray}
G_{0d}^{A}=YD_{0}^{A}\alpha^{A}G^{A}_{dd},\label{G0dA}
\end{eqnarray}
where $D_{0}^{A}$ is the reference site Green function when the
upper arm of the ring is cut off,
\begin{eqnarray}
D_{0}^{A}=\frac{1}{\omega -i\eta
-\epsilon_{0}-\alpha^{A}(i_{\ell}^{2}+i_{r}^{2})},\label{D0}
\end{eqnarray}
and $Y$ denotes the interference coupling,
\begin{eqnarray}
Y=i_{\ell}j_{\ell}e^{-i\phi_{\ell}}+i_{r}j_{r}e^{i\phi_{r}}.
\label{Y}
\end{eqnarray}
The retarded Green function $G_{0d}^{R}$ is obtained from Eq.
(\ref{G0dA}) by interchanging  `$A$' into `$R$'. A similar
calculation yields
\begin{eqnarray}
G_{d0}^{A}=Y^{\ast}D_{0}^{A}\alpha^{A}G^{A}_{dd}.\label{Gd0A}
\end{eqnarray}
[We remind the reader that the two leads are assumed to be
identical except that they are connected to electronic reservoirs
with different chemical potentials.] Turning now to the
calculation of $G^{<}_{0d}$ and $G_{d0}^{<}$,  we first apply the
rule (\ref{rule}) to Eq. (\ref{G0d}), to obtain
\begin{eqnarray}
G^{<}_{0d}&=&g_{0}^{R}\Bigl [\sum_{k}v^{\ast}_{k}\Bigl
(V_{k}(g_{k}^{R}G^{<}_{dd}+g_{k}^{<}G^{A}_{dd})\nonumber\\
&+&v_{k}(g_{k}^{R}G_{0d}^{<}+g_{k}^{<}G_{0d}^{A})\Bigr
)+\{k\rightarrow p\}\Bigr ].
\end{eqnarray}
Then we collect the coefficients of $G_{0d}^{<}$ [using Eq.
(\ref{D0})] to find
\begin{eqnarray}
G^{<}_{0d}&=&D_{0}^{R}\Bigl [\sum_{k}v^{\ast}_{k}\Bigl
(V_{k}(g_{k}^{R}G^{<}_{dd}+g_{k}^{<}G^{A}_{dd})\nonumber\\
&+&v_{k}g_{k}^{<}G_{0d}^{A}\Bigr )+\{k\rightarrow p\}\Bigr ].
\end{eqnarray}
Finally we insert here Eq. (\ref{G0dA}) to obtain
\begin{eqnarray}
G^{<}_{0d}&=&\alpha^{R}D_{0}^{R}Y(
G^{<}_{dd}-f_{\ell}G^{A}_{dd})+f_{\ell}\alpha^{A}D^{A}_{0}YG^{A}_{dd}\nonumber\\
&+&\Delta D^{R}_{0}G^{A}_{dd}(f_{r}-f_{\ell})i_{r}(J_{r}^{R}(\Phi
))^{\ast}e^{i\phi_{r}}.\label{G0dK}
\end{eqnarray}
A similar calculation yields
\begin{eqnarray}
G^{<}_{d0}&=&\alpha^{A}D_{0}^{A}Y^{\ast}(
G^{<}_{dd}+f_{\ell}G^{R}_{dd})-f_{\ell}\alpha^{R}D^{R}_{0}Y^{\ast}G^{R}_{dd}\nonumber\\
&+&\Delta D^{A}_{0}G^{R}_{dd}(f_{r}-f_{\ell})i_{r}J_{r}^{R}(\Phi
)e^{-i\phi_{r}}.\label{Gd0K}
\end{eqnarray}
Here we have introduced the effective couplings connecting the
quantum dot to the right part of the ring,
\begin{eqnarray}
J_{r}^{R}(\Phi
)=j_{r}+i_{r}\alpha^{R}D^{R}_{0}(i_{\ell}j_{\ell}e^{i\Phi}+i_{r}j_{r}),\label{JJr}
\end{eqnarray}
and to the left side,
\begin{eqnarray}
J_{\ell}^{R}(\Phi
)=j_{\ell}+i_{\ell}\alpha^{R}D^{R}_{0}(i_{\ell}j_{\ell}+i_{r}j_{r}e^{-i\Phi}),
\label{JJl}
\end{eqnarray}
and used the relation
\begin{eqnarray}
D^{R}_{0}-D^{A}_{0}= -\Delta
D^{R}_{0}D^{A}_{0}(i_{\ell}^{2}+i_{r}^{2}).
\end{eqnarray}

Introducing all these results into  $I_{1}$, Eq. (\ref{cur1}),
gives that partial current in terms of the dot Green function,
\begin{eqnarray}
&&I_{1}=e\int\frac{d\omega}{2\pi}\Bigl \{2i\sin\Phi
(i_{\ell}j_{\ell}i_{r}j_{r})\Bigl
[(\alpha^{A})^{2}D^{A}_{0}G^{A}_{dd}-cc\Bigr
]f_{\ell}\nonumber\\
&+& \Bigl [\alpha^{R}j_{\ell}J^{R}_{\ell}(-\Phi )-cc\Bigr ]\Bigl
[G^{<}_{dd}+f_{\ell}(G^{R}_{dd}-G^{A}_{dd})\Bigr ]\nonumber\\
&+&\Bigl [(J_{r}^{R}(\Phi
))^{\ast}j_{\ell}i_{\ell}i_{r}e^{i\Phi}\alpha^{R}D^{R}_{0}G^{A}_{dd}-cc\Bigr
]\Delta (f_{r}-f_{\ell})\Bigr \}.
\nonumber\\
 \label{I1}
\end{eqnarray}
[Note that $\Delta^{\ast}=-\Delta$.]  Examining the expression for
the partial current $I_{3}$, Eq. (\ref{currents}), it is seen that
it is obtained from  $I_{1}$, upon the replacements
$\ell\leftrightarrow r$ with $k\leftrightarrow p$, and
$\phi_{\ell}\leftrightarrow -\phi_{r}$, namely $\Phi\rightarrow
-\Phi$. Then [see Eqs. (\ref{JJr}) and (\ref{JJl})]
$J^{R}_{\ell}(-\Phi )\leftrightarrow J^{R}_{r}(\Phi )$.

Next we consider the partial current $I_{2}$. A similar
calculation to the one leading to Eq. (\ref{cur1}) yields
\begin{eqnarray}
I_{2}&=&e\int\frac{d\omega}{2\pi}\Bigl (-\Delta i_{\ell}^{2}\Bigl
[G^{<}_{00}+f_{\ell}(G^{R}_{00}-G^{A}_{00})\Bigr ]\nonumber\\
&+&j_{\ell}i_{\ell}e^{-i\phi_{\ell}}\Bigl [
\alpha^{R}G^{<}_{d0}+\Delta f_{\ell}G^{A}_{d0}\Bigr
]\nonumber\\&-&j_{\ell}i_{\ell}e^{i\phi_{\ell}}\Bigl [\Delta
f_{\ell}G^{R}_{0d}+\alpha^{A}G^{<}_{0d}\Bigr ]\Bigr ).\label{cur2}
\end{eqnarray}
In order to express this current in terms of the dot Green
function, we need  to calculate the reference site Green function
$G_{00}$. The equation-of-motion for the time-ordered counterpart
gives
\begin{eqnarray}
G_{00}=g_{0}+g_{0}\Bigl [\sum_{k}v_{k}^{\ast}G_{k0}+\{
k\leftrightarrow p\}\Bigr ],\label{G00}
\end{eqnarray}
with
\begin{eqnarray}
G_{k0}=V_{k}g_{k}G_{d0}+v_{k}g_{k}G_{00}.\label{Gk0}
\end{eqnarray}
Making use of Eqs. (\ref{D0}) and (\ref{Gd0A}), we  find
\begin{eqnarray}
&&G^{R}_{00}=D^{R}_{0} +(D^{R}_{0}\alpha^{R})^{2}|Y|^{2}
G^{R}_{dd}.\label{G00R}
\end{eqnarray}
The advanced function $G^{A}_{00}$ is given by this equation upon
interchanging `$R$' with `$A$'. Next we apply the rule
(\ref{rule}) to Eqs. (\ref{G00}) and (\ref{Gk0}) to obtain
\begin{eqnarray}
G^{<}_{00}&=&D^{R}_{0}\Bigl [Y\alpha^{R}G^{<}_{d0} +\Delta
(i_{\ell}^{2}f_{\ell}+i_{r}^{2}f_{r})G^{A}_{00}\nonumber\\
&+&\Delta
(i_{\ell}j_{\ell}e^{-i\phi_{\ell}}f_{\ell}+i_{r}j_{r}e^{i\phi_{r}}f_{r})G^{A}_{d0}
\Bigr ].
\end{eqnarray}
It remains to insert here Eqs. (\ref{Gd0A}), (\ref{Gd0K}), and
(\ref{G00R}), to obtain
\begin{eqnarray}
&&G_{00}^{<}=\Delta
D^{R}_{0}D^{A}_{0}(i_{\ell}^{2}f_{\ell}+i_{r}^{2}f_{r})
\nonumber\\
&+&|Y\alpha^{R}D_{0}^{R}|^{2}\Bigl
[G^{<}_{dd}+f_{\ell}(G^{R}_{dd}-G^{A}_{dd})\Bigr
]\nonumber\\
&+&\Bigl [\Delta \alpha^{R}D^{R}_{0}D^{A}_{0}Yi_{r}J_{r}^{R}(\Phi
)e^{-i\phi_{r}}  (f_{r}-f_{\ell})G^{R}_{dd} -cc\Bigr
]\nonumber\\
&+&|Y|^{2}f_{\ell}\Bigl
[(\alpha^{A}D_{0}^{A})^{2}G^{A}_{dd}-cc\Bigr ].\label{G00K}
\end{eqnarray}
The first term here is the contribution of the lower arm of the
ring alone; The other terms arise from interference.

Introducing these results into Eq. (\ref{cur2}) for $I_{2}$, we
find
\begin{eqnarray}
&&I_{2}=e\int\frac{d\omega}{2\pi}\Bigl
\{D^{R}_{0}D^{A}_{0}i_{\ell}^{2}i_{r}^{2}\Delta^{2}(f_{\ell}-f_{r})\nonumber\\
&-&2i\sin\Phi (i_{\ell}j_{\ell}i_{r}j_{r})\Bigl
[(\alpha^{A})^{2}D^{A}_{0}G^{A}_{dd}-cc\Bigr
]f_{\ell}\nonumber\\
&+&\Bigl [\alpha^{R}J^{R}_{\ell}(-\Phi )\Bigl ((J_{\ell}^{R}(-\Phi
))^{\ast}-j_{\ell}\Bigr )-cc \Bigr ]
\nonumber\\
&&\times \Bigl [G^{<}_{dd}+f_{\ell}(G^{R}_{dd}-G^{A}_{dd})\Bigr ]
\nonumber\\
&+&\Bigl
[\alpha^{R}D^{A}_{0}i_{\ell}i_{r}e^{-i\Phi}G^{R}_{dd}J_{r}^{R}(\Phi
)\nonumber\\
&&\times\Bigl (j_{\ell} -\Delta
D^{R}_{0}i_{\ell}e^{i\phi_{\ell}}Y\Bigr )- cc\Bigr ]\Delta
(f_{r}-f_{\ell})\Bigr \}. \label{I2}
\end{eqnarray}
Examining the expression for the partial current $I_{4}$, Eq.
(\ref{currents}), it is seen that it is obtained from  $I_{2}$,
upon the replacements $\ell\leftrightarrow r$ and
$\phi_{\ell}\leftrightarrow -\phi_{r}$.

\subsection{Current conservation}

Having obtained the partial currents in terms of the dot Green
functions, we now examine the consequences. The important point to
bear in mind is that in the presence of interactions (confined to
the quantum dot alone), those Green functions are not known, and
may be found only approximately. Therefore, imposing current
conservation will yield  general relations which the $G_{dd}$'s
have to satisfy.

Current conservation means (see Fig. \ref{fig1}) that
$I_{1}+I_{3}=0$, and $I_{2}+I_{4}=0$. A lengthy calculation of the
sum $I_{2}+I_{4}$, using Eq. (\ref{I2}), shows that it indeed
vanishes. In contrast, the sum of the currents on the
interferometer arm containing the dot, using Eq. (\ref{I1}), gives
\begin{eqnarray}
&&I_{1}+I_{3} =e\int\frac{d\omega}{2\pi}\nonumber\\
&&\times\Bigl [(\Sigma^{R}_{\rm ext}-\Sigma^{A}_{\rm
ext})G_{dd}^{<}+\Sigma^{<}_{\rm ext}(G_{dd}^{A}-G_{dd}^{R})\Bigr
],\label{conv}
\end{eqnarray}
in which $\Sigma_{\rm ext}$ denotes the self-energy of the dot
Green function, which arises from the connection of the dot to the
interferometer and the leads. This relation may be verified as
follows. The external self-energy, $\Sigma_{\rm ext}$, is found
from the Dyson equation (\ref{Dyson}), using only the
non-interacting parts of the Hamiltonian, Eq. (\ref{ham}), and Eq.
(\ref{Gdk}),
\begin{eqnarray}
\Sigma_{\rm ext}G_{dd}&=&\Bigl [\sum_{k}V_{k}g_{k}\Bigl
(V_{k}G_{dd}+v_{k}G_{0d}\Bigr )+\{k\rightarrow p\}\Bigr
].\nonumber\\
\end{eqnarray}
Employing Eqs. (\ref{G0dA}) and (\ref{G0dK}), together with the
rule (\ref{rule}), one then finds
\begin{eqnarray}
&&\Sigma_{\rm ext}^{R}=\alpha^{R}\Bigl (j_{\ell}^{2}+j_{r}^{2}
+\alpha^{R}D_{0}^{R}|Y|^{2}\Bigr ),\label{sigmaext}
\end{eqnarray}
with an analogous expression for $\Sigma_{\rm ext}^{A}$, and
\begin{eqnarray}
\Sigma_{\rm ext}^{<}&=&\Delta \Bigl [f_{\ell}|J_{\ell}^{R}(\Phi
)|^{2}+f_{r}|J_{r}^{R}(\Phi )|^{2}\Bigr ].\label{sigmaextK}
\end{eqnarray}
When one combines $I_{1}$ and $I_{3}$, and uses the results Eqs.
(\ref{sigmaext}) and (\ref{sigmaextK}), one arrives at Eq.
(\ref{conv}). Note that, using Eqs. (\ref{JJr}), (\ref{JJl}) and
(\ref{sigmaext}), one has
\begin{eqnarray}
\Sigma^{A}_{\rm ext}-\Sigma^{R}_{\rm ext}&=&\Delta \Bigl
[|J_{\ell}^{R}(\Phi )|^{2}+|J_{r}^{R}(\Phi )|^{2}\Bigr
]\nonumber\\
&=&\Delta \Bigl [|J_{\ell}^{R}(-\Phi )|^{2}+|J_{r}^{R}(-\Phi
)|^{2}\Bigr ].\label{relationsig}
\end{eqnarray}

When the electronic system is un-biased, namely, when the chemical
potentials on both reservoirs are identical, we have
\begin{eqnarray}
f_{\ell}=f_{r}\equiv f_{th}.\label{thermal}
\end{eqnarray}
Then Eqs. (\ref{sigmaextK}) and (\ref{relationsig}) give
\begin{eqnarray}
\Sigma_{\rm ext}^{<}=f_{th}(\Sigma_{\rm ext}^{A}-\Sigma_{\rm
ext}^{R}).
\end{eqnarray}
Without the bias,   one also has \cite{langreth}
\begin{eqnarray}
G^{<}_{dd}=f_{th}(G^{A}_{dd}-G^{R}_{dd}).
\end{eqnarray}
It follows that without a bias, the integrand in Eq. (\ref{conv})
vanishes. In other words, when the ring is not biased, current
conservation is trivially satisfied.

Another case in which Eq. (\ref{conv}) is trivially satisfied is
when the dot is free of any interactions. Then the dot Green
function (marked by the superscript `0') obeys
\begin{eqnarray}
G^{R0}_{dd}&=&\frac{1}{\omega -\epsilon_{d}-\Sigma^{R}_{\rm
ext}},\nonumber\\
G^{R0}_{dd}-G^{A0}_{dd}&=&G^{A0}_{dd}\Bigl [\Sigma_{\rm
ext}^{R}-\Sigma_{\rm ext}^{A}\Bigr ]G^{R0}_{dd},\label{Gdd0}
\end{eqnarray}
where for simplicity it has been assumed that there is only a
single electronic level on the dot, denoted $\epsilon_{d}$. For
the non-interacting system one  also has \cite{langreth}
\begin{eqnarray}
G^{<0}_{dd}=G^{A0}_{dd}\Sigma_{\rm ext}^{<}G^{R0}_{dd},
\end{eqnarray}
and therefore, again, the integrand in Eq. (\ref{conv}) vanishes.

Had we known the exact forms of $G_{dd}^{R,A}$ and $G^{<}_{dd}$
for the interacting electronic system, we would have found that
current conservation  is also satisfied when the ring is biased.
However, as  mentioned above, the dot Green function is not known
exactly. Therefore, we may regard the relation Eq. (\ref{conv}) as
a condition imposed on $G_{dd}^{R,A}$ and $G_{dd}^{<}$. In order
to make a practical use of this condition, we  assume that  the
main contribution to the $\omega$-integration in Eq. (\ref{conv})
comes from frequencies at about the Fermi level of the electrons
(remembering that the ring is only slightly biased), and therefore
we may write
\begin{eqnarray}
G^{<}_{dd}=\Sigma_{\rm
ext}^{<}\frac{G^{R}_{dd}-G^{A}_{dd}}{\Sigma_{\rm
ext}^{R}-\Sigma_{\rm ext}^{A}}.\label{GddK}
\end{eqnarray}
This approximation  is insufficient to determine the dot Green
function, but at least it eliminates the necessity to calculate
the Keldysh Green function $G_{dd}^{<}$, and ensures that the
current through the ring is conserved. Without a bias voltage, Eq.
(\ref{GddK}) holds exactly, yielding in particular the charge in
the dot, which is equal to the expectation value of the dot
occupation, $ n_{d} =-i\int(d\omega /2\pi )G^{<}_{dd}(\omega )$,
[see Eq. (\ref{KELDYSH})]. With a finite bias, Eq. (\ref{GddK}) is
only approximate. However, the implied dependence of $ n_{d} $ on
the bias voltage will still obey current conservation. The
approximation leading to Eq. (\ref{GddK}) is sometimes referred to
as the `wide-band' approximation. \cite{jauho}

With the approximation (\ref{GddK}) and Eqs. (\ref{sigmaextK}) and
(\ref{relationsig}), one has
\begin{eqnarray}
G^{<}_{dd}+f_{\ell}(G^{R}_{dd}-G^{A}_{dd})&=&\frac{G^{R}_{dd}-G^{A}_{dd}}{\Sigma^{R}_{\rm
ext}-\Sigma^{A}_{\rm ext}}\Delta |J_{r}^{R}(\Phi )|^{2}
(f_{r}-f_{\ell}),\nonumber\\
G^{<}_{dd}+f_{r}(G^{R}_{dd}-G^{A}_{dd})&=&\frac{G^{R}_{dd}-G^{A}_{dd}}{\Sigma^{R}_{\rm
ext}-\Sigma^{A}_{\rm ext}}\Delta |J_{\ell}^{R}(\Phi )|^{2}
(f_{\ell}-f_{r}).\nonumber\\
\label{combi}
\end{eqnarray}
We emphasize again that when the ring is not biased, or when the
dot is free of any interactions, the relations (\ref{GddK}) and
(\ref{combi}) are always satisfied.

\subsection{The current through the ring}

A glance at Fig. \ref{fig1} shows that the current through the
ring, $I$, is given by $I=I_{1}+I_{2}=-I_{3}-I_{4}$. This current
is conveniently found from Eqs. (\ref{I1}) and (\ref{I2}) by
calculating $(I_{1}+I_{2}-I_{3}-I_{4})/2$. The terms proportional
to $\sin\Phi$ are then cancelled, and one is left with
\begin{eqnarray}
I&=&e\int\frac{d\omega}{2\pi}\Bigl \{ -\frac{\Delta}{2}
|J_{\ell}^{R}(-\Phi )|^{2}\Bigl
[G^{<}_{dd}+f_{\ell}(G^{R}_{dd}-G^{A}_{dd})\Bigr ]\nonumber\\
&+&\frac{\Delta}{2} |J_{r}^{R}(-\Phi )|^{2}\Bigl
[G^{<}_{dd}+f_{r}(G^{R}_{dd}-G^{A}_{dd})\Bigr ]\nonumber\\
&+&(f_{\ell}-f_{r})\Delta^{2}i_{\ell}^{2}i_{r}^{2}|D^{R}_{0}|^{2}\Bigl
[1+G^{R}_{dd}\Sigma^{R}_{\rm ext}+G^{A}_{dd}\Sigma^{A}_{\rm
ext}\Bigr ]\nonumber\\
 &+&(f_{\ell}-f_{r})\Delta^{2}i_{\ell}i_{r}j_{\ell}j_{r}\cos\Phi\nonumber\\
 &&\times D^{R}_{0}\Bigl
 [(1+\alpha^{A}D^{A}_{0}(i_{\ell}^{2}+i_{r}^{2})\Bigr ]
  (G^{R}_{dd}+G^{A}_{dd})\Bigr \}.\label{iI}
 \end{eqnarray}
[Note that the quantity $D^{R}_{0}
[1+\alpha^{A}D^{A}_{0}(i_{\ell}^{2}+i_{r}^{2}) ]$ is real, see Eq.
(\ref{D0}).] Making use of the approximation (\ref{GddK}) and the
resulting relations (\ref{combi}), the current through the ring
takes the form
\begin{eqnarray}
I=I_{\rm ref}+I_{\rm dot}+I_{\rm int},\label{totalI}
\end{eqnarray}
where the first term here, $I_{\rm ref}$, reduces to the current
through the reference arm when the other arm is disconnected,
\begin{eqnarray}
&&I_{\rm ref}=e\int
\frac{d\omega}{2\pi}(f_{\ell}-f_{r})\Delta^{2}i_{\ell}^{2}i_{r}^{2}|D^{R}_{0}|^{2}
\Bigl (1+G^{R}_{dd}\Sigma^{R}_{\rm ext} \nonumber\\
&+& G^{A}_{dd}\Sigma^{A}_{\rm ext}+\Sigma^{R}_{\rm
ext}\Sigma^{A}_{\rm
ext}\frac{G^{R}_{dd}-G^{A}_{dd}}{\Sigma^{R}_{\rm
ext}-\Sigma^{A}_{\rm ext}}\Bigr ).\label{Iref}
\end{eqnarray}
  Similarly, the current
$I_{\rm dot}$, which reduces to the one flowing in the absence of
the reference arm, is
\begin{eqnarray}
I_{\rm dot}&=&e\int
\frac{d\omega}{2\pi}(f_{\ell}-f_{r})\Delta^{2}j_{\ell}^{2}j_{r}^{2}
\nonumber\\
&&\times |1+\alpha^{R}D^{R}_{0}(i_{\ell}^{2}+i_{r}^{2})|^{2}
\frac{G^{R}_{dd}-G^{A}_{dd}}{\Sigma^{R}_{\rm ext}-\Sigma^{A}_{\rm
ext}}.\label{Idot}
\end{eqnarray}
Each of these currents is `dressed' by processes in which the
electrons travel through the other branch. As might be expected,
the interference between the two branches always appears via the
product $j_{\ell}j_{r}i_{\ell}i_{r}\cos\Phi$. In addition to
appearing implicitly, via $\Sigma_{\rm ext}$, in $I_{\rm dot}$ and
$I_{\rm ref}$, this product appears explicitly in the last member
in Eq. (\ref{totalI})
\begin{eqnarray}
&&I_{\rm int}=e\int
\frac{d\omega}{2\pi}(f_{\ell}-f_{r})\Delta^{2}i_{\ell}i_{r}j_{\ell}j_{r}D^{R}_{0}
 [1+\alpha^{A}D^{A}_{0}(i_{\ell}^{2}+i_{r}^{2})]
\nonumber\\
&&\times\cos\Phi \Bigl (G^{R}_{dd}+G^{A}_{dd}+(\Sigma^{R}_{\rm
ext}+\Sigma^{A}_{\rm
ext})\frac{G^{R}_{dd}-G^{A}_{dd}}{\Sigma^{R}_{\rm
ext}-\Sigma^{A}_{\rm ext}}\Bigr ).\label{Iint}
\end{eqnarray}

An important aspect of the result for the current through the
interferometer, Eq. (\ref{totalI}), is that it is an {\it even}
function of the flux $\Phi$, since both $G^{R,A}_{dd}$ and
$\Sigma^{R,A}_{\rm ext}$ are even functions of $\Phi$. Namely, the
current through the interferometer obeys the Onsager
relations.\cite{buttions} It is interesting to note that this
property is not apparent from Eq. (\ref{iI}); However, once  we
use  the relation (\ref{GddK}), which ensures current
conservation, then the flux-parity of $I$ becomes clear.

To present the current in a more transparent manner, we write the
couplings $i_{\ell}$, $i_{r}$, $j_{\ell}$, and $j_{r}$, in terms
of the partial widths they induce on the localized levels of the
interferometer  (the one on the reference arm and the one on the
dot). In general, when a localized level is coupled by a matrix
element $u_{k}$ to a continuum of states of energies
$\epsilon_{k}$, it becomes a resonance of width
\begin{eqnarray}
\Gamma (\omega )=\pi\sum_{k}|u_{k}|^{2}\delta (\omega
-\epsilon_{k}).
\end{eqnarray}
Making use of the matrix elements, Eqs. (\ref{coupling}), in
conjunction with Eq. (\ref{delta}), we define
\begin{eqnarray}
\gamma_{\ell }=i_{\ell}^{2}\frac{\Delta}{2i},\ \ \gamma_{r
}=i_{r}^{2}\frac{\Delta}{2i},\label{gammal}
\end{eqnarray}
for the partial widths on the reference site, and
\begin{eqnarray}
\Gamma_{\ell }=j_{\ell}^{2}\frac{\Delta}{2i},\ \ \Gamma_{r
}=j_{r}^{2}\frac{\Delta}{2i},\label{GAMMAL}
\end{eqnarray}
for the partial widths on the quantum dot. In accordance with the
wide-band approximation used to obtain the current Eq.
(\ref{totalI}), we  also neglect the frequency-dependence of those
widths. As a result, the various parameters appearing in Eqs.
(\ref{totalI}), (\ref{Iref}), (\ref{Idot}), and (\ref{Iint}) can
be put in the following forms. Firstly, we consider the prefactor
in the expression for $I_{\rm ref}$,
\begin{eqnarray}
\Delta^{2}|i_{\ell}^{2}i_{r}^{2}D^{R}_{0}|^{2}=
-2i\frac{\gamma_{\ell}\gamma_{r}}{\gamma_{\ell}+\gamma_{r}}(D^{R}_{0}-D^{A}_{0})
\equiv -T_{B},\label{Irefco}
\end{eqnarray}
where $T_{B}$ is the transmission coefficient of the reference arm
of the interferometer (when decoupled from the quantum dot). In
the wide-band approximation (in which the energy is taken to be at
the middle of the band)
\begin{eqnarray}
T_{B}=\frac{4\gamma_{\ell}\gamma_{r}}{\epsilon^{2}_{0}+(\gamma_{\ell}+\gamma_{r})^{2}}.
\label{TB}
\end{eqnarray}
Secondly, the prefactor in the expression for $I_{\rm dot}$
becomes
\begin{eqnarray}
&&\Delta^{2}j_{\ell}^{2}j_{r}^{2}|1+\alpha^{R}D_{0}^{R}
(i_{\ell}^{2}+i_{r}^{2})|^{2} \equiv
-4\Gamma_{\ell}\Gamma_{r}X_{B},\label{Idotco}
\end{eqnarray}
with $X_{B}$ given in Eq. (\ref{XB}).
%\begin{eqnarray}
%X_{B}=1-T_{B}\frac{(\gamma_{\ell}+\gamma_{r})^{2}}{4\gamma_{\ell}\gamma_{r}}.
%\end{eqnarray}
Finally, the coefficient in  $I_{\rm int}$ is
\begin{eqnarray}
&&\Delta^{2}i_{\ell}i_{r}j_{\ell}j_{r}D^{R}_{0} \Bigl
(1+\alpha^{A}D_{0}^{A}(i_{\ell}^{2}+i_{r}^{2})\Bigr )
\nonumber\\
&=&-2sgn(\epsilon_{0})\sqrt{T_{B}\Gamma_{\ell}\Gamma_{r}X_{B}
 }.\label{Iintco}
\end{eqnarray}

For the sake of completeness, we add here the `external'
self-energy of the dot Green function, expressed in terms of the
partial resonance widths,
\begin{eqnarray}
\Sigma^{R}_{\rm ext}&=&-i\Bigl (\Gamma_{\ell}+\Gamma_{r}
-\frac{T_{B}(\gamma_{\ell}+\gamma_{r})}{4}Z_{B} \Bigr )
\nonumber\\
&+&sgn(\epsilon_{0})\frac{Z_{B}}{2}\sqrt{T_{B}\gamma_{\ell}\gamma_{r}X_{B}},
\label{sigmaRWB}
\end{eqnarray}
with
\begin{eqnarray}
Z_{B}=\frac{\Gamma_{\ell}}{\gamma_{r}}+\frac{\Gamma_{r}}{\gamma_{\ell}}-2\cos\Phi
\sqrt{\frac{\Gamma_{\ell}\Gamma_{r}}{\gamma_{\ell}\gamma_{r}}}.
\label{ZB}
\end{eqnarray}
It is thus seen that both the imaginary and the real parts of
$\Sigma_{\rm ext}$ depend on the flux threading the
interferometer, through the interference term $Z_{B}$. This
expression for the external self-energy differs from the one
reported in Ref. \onlinecite{hofstetter}, in which the imaginary
part of $\Sigma^{R}_{\rm ext}$ is independent of the flux, while
its real part vanishes for $\Phi =\pi /2$. Although the details of
$\Sigma_{\rm ext}^{R}$ are necessarily model-dependent, the result
given there, which apparently neglects any scattering on the
reference arm, is obviously rather restricted to a very specific
situation.

Inserting the results (\ref{Irefco}), (\ref{Idotco}), and
(\ref{Iintco}) into Eqs. (\ref{Iref}), (\ref{Idot}), and
(\ref{Iint}) yields our final result for the current through the
interferometer, Eq. (\ref{ITRAN}),
%\begin{eqnarray}
%I&=&e\int\frac{d\omega}{2\pi}(f_{r}-f_{\ell}) \Bigl \{T_{B}\Bigl
%(1+G^{R}_{dd}\Sigma^{R}_{\rm
%ext}\nonumber\\
%&+&G^{A}_{dd}\Sigma^{A}_{\rm ext}+\Sigma^{R}_{\rm
%ext}\Sigma^{A}_{\rm
%ext}\frac{G^{R}_{dd}-G^{A}_{dd}}{\Sigma^{R}_{\rm
%ext}-\Sigma^{A}_{\rm ext}}\Bigr )\nonumber\\
%&+&4\Gamma_{\ell}\Gamma_{r}X_{B}\frac{G^{R}_{dd}-G^{A}_{dd}}{\Sigma^{R}_{\rm
%ext}-\Sigma^{A}_{\rm ext}}
%+\sqrt{T_{B}\Gamma_{\ell}\Gamma_{r}X_{B}}2\cos\Phi\nonumber\\
%&&\times\Bigl ( G^{R}_{dd}+G^{A}_{dd}+(\Sigma^{R}_{\rm
%ext}+\Sigma^{A}_{\rm
%ext})\frac{G^{R}_{dd}-G^{A}_{dd}}{\Sigma^{R}_{\rm
%ext}-\Sigma^{A}_{\rm ext}}\Bigr )\Bigr \},\label{Isofi}
%\end{eqnarray}
where for simplicity we have chosen the sign of the on-site energy
on the reference site to be positive. We note that this result is
not the same as the ones given in Refs. \onlinecite{bulka} and
\onlinecite{hofstetter},  which neglected the scattering on the
reference site. On the other hand, our expression reduces to the
result obtained from a straightforward calculation (that does not
employ the Keldysh technique), for an interaction-free system, as
will be shown below.

When there are no interactions on the dot, then using Eqs.
(\ref{Gdd0}), and denoting the interaction-free current by
$I^{0}$, one has
\begin{eqnarray}
I^{0}&=&e\int\frac{d\omega}{2\pi}(f_{r}-f_{\ell})\Bigl \{ T_{B}
|1+G^{R0}_{dd}\Sigma^{R}_{\rm ext}|^{2}\nonumber\\
&+&4\Gamma_{\ell}\Gamma_{r}X_{B}|G^{R0}_{dd} |^{2}
+2\sqrt{T_{B}\Gamma_{\ell}\Gamma_{r}X_{B}}\cos\Phi \nonumber\\
&&\times \Bigl (G^{R0}_{dd}(1+\Sigma^{A}_{\rm
ext}G^{A0}_{dd})+cc\Bigr ) \Bigr \}. \label{Inint}
\end{eqnarray}
Noting now that
\begin{eqnarray}
(G^{R0}_{dd})^{-1}+\Sigma^{R}_{\rm ext}\equiv ({\cal G}^{0})^{-1}
\equiv\omega -\epsilon_{d}\label{g00}
\end{eqnarray}
is a  {\it real} function, $I^{0}$ can be written as
\begin{eqnarray}
I^{0}&=&
e\int\frac{d\omega}{2\pi}(f_{r}-f_{\ell})|G^{R0}_{dd}|^{2}\nonumber\\
&&\times|\sqrt{T_{B}}({\cal
G}^{0})^{-1}e^{i\Phi}+2\sqrt{\Gamma_{\ell}\Gamma_{r}X_{B}}|^{2}.
\label{2slit}
\end{eqnarray}
This result reproduces the one found using scattering-matrix
description. \cite{avi}

\subsection{The circulating current}

In order to calculate the current circulating {\it around} the
interferometer, we consider the quantity
$(I_{1}-I_{2}-I_{3}+I_{4})/2$ employing Eqs. (\ref{I1}) and
(\ref{I2}), and then take its antisymmetric part with respect to
the flux. Clearly the first term in Eq. (\ref{I2}) will eventually
disappear, since it is independent of $\Phi$. Therefore, we will
not include that term. We have
\begin{eqnarray}
&&\frac{I_{1}-I_{2}-I_{3}+I_{4}}{2}\nonumber\\
&=&e\int\frac{d\omega}{2\pi}\Bigl
\{2i\frac{f_{\ell}+f_{r}}{2}\Bigl [\frac{\partial\Sigma^{A}_{\rm
ext}}{\partial\Phi}G^{A}_{dd}-cc\Bigr ]\nonumber\\
&+&\frac{1}{2}\Bigl [\alpha^{R}J^{R}_{\ell}(-\Phi )\Bigl
(2j_{\ell}-(J^{R}_{\ell}(-\Phi ))^{\ast}\Bigr )-cc\Bigr
]\nonumber\\
&&\times\Bigl [G^{<}_{dd}+f_{\ell}(G^{R}_{dd}-G^{A}_{dd})\Bigr
]\nonumber\\
&-&\frac{1}{2}\Bigl [\alpha^{R}J^{R}_{r}(-\Phi )\Bigl
(2j_{r}-(J^{R}_{r}(-\Phi ))^{\ast}\Bigr )-cc\Bigr
]\nonumber\\
&&\times\Bigl [G^{<}_{dd}+f_{r}(G^{R}_{dd}-G^{A}_{dd})\Bigr
]\nonumber\\
&+&(f_{r}-f_{\ell})\Delta\Bigl [G^{R}_{dd}\Bigl
(i_{\ell}^{2}i_{r}^{2} D^{R}_{0}D^{A}_{0}(\Delta\Sigma^{R}_{\rm
ext} -\alpha^{R}\alpha^{A}(j_{\ell}^{2}+j_{r}^{2}))\nonumber\\
&-&i_{\ell}i_{r}j_{\ell}j_{r}\cos\Phi
(\alpha^{A}D_{0}^{A}+\alpha^{R}D^{R}_{0})\nonumber\\
&-&|\alpha^{R}D^{R}_{0}|^{2}i_{\ell}i_{r}j_{\ell}j_{r}(i_{r}^{2}e^{-i\Phi}+i_{\ell}^{2}e^{i\Phi})
\Bigr )-cc\Bigr ]\Bigr \},
\end{eqnarray}
where in order to obtain the first term here we have used Eq.
(\ref{sigmaext}). Clearly, when the system is unbiased and the
distribution functions $f_{\ell}$ and $f_{r}$ are the thermal ones
[see Eq. (\ref{thermal})], only that term survives. In order to
explore the situation in which the system is biased, we proceed as
follows. Firstly, we consider the contribution of the last square
brackets, multiplying the difference $f_{r}-f_{\ell}$. The
antisymmetric part with respect to $\Phi$ is just
\begin{eqnarray}
i\sin\Phi
G^{R}_{dd}i_{\ell}i_{r}j_{\ell}j_{r}|\alpha^{R}D^{R}_{0}|^{2}(i_{r}^{2}-i_{\ell}^{2})
-cc.\label{per1}
\end{eqnarray}
Next we consider the two terms involving the combinations
$G^{<}_{dd}+f_{\ell}(G^{R}_{dd}-G^{A}_{dd})$ and
$G^{<}_{dd}+f_{r}(G^{R}_{dd}-G^{A}_{dd})$, using Eqs.
(\ref{combi}). The sum of these two terms becomes
\begin{eqnarray}
&&\frac{\Delta}{2}(f_{r}-f_{\ell})\frac{G^{R}_{dd}-G^{A}_{dd}}{\Sigma^{R}_{\rm
ext}-\Sigma^{A}_{\rm ext}}\Bigl [|J_{r}^{R}(\Phi )|^{2}\Bigl
(\Delta |J_{\ell}^{R}(-\Phi
)|^{2}\nonumber\\
&+&2\alpha^{R}j_{\ell}J^{R}_{\ell}(-\Phi
)-2\alpha^{A}j_{\ell}(J^{R}_{\ell}(-\Phi ))^{\ast}\Bigr
)\nonumber\\
&+&|J_{\ell}^{R}(\Phi )|^{2}\Bigl (\Delta |J_{r}^{R}(-\Phi
)|^{2}\nonumber\\
&+&2\alpha^{R}j_{r}J^{R}_{r}(-\Phi
)-2\alpha^{A}j_{r}(J^{R}_{r}(-\Phi ))^{\ast}\Bigr )\Bigr ].
\end{eqnarray}
The antisymmetric part with respect to $\Phi$ is
\begin{eqnarray}
&&\frac{\Delta}{2}(f_{r}-f_{\ell})\frac{G^{R}_{dd}-G^{A}_{dd}}{\Sigma^{R}_{\rm
ext}-\Sigma^{A}_{\rm
ext}}i_{\ell}i_{r}j_{\ell}j_{r}2i\sin\Phi\nonumber\\
&&\times\alpha^{R}\alpha^{A}\Bigl
[D^{R}_{0}D^{A}_{0}(\Sigma^{R}_{\rm
ext}+\Sigma^{A}_{\rm ext})(i_{r}^{2}-i_{\ell}^{2}) \nonumber\\
&+&2D^{A}_{0}(1+\alpha^{R}D^{R}_{0}(i_{\ell}^{2}+i_{r}^{2}))(j_{r}^{2}-j_{\ell}^{2})\Bigr
].\label{per2}
\end{eqnarray}
Inspecting Eqs. (\ref{per1}) and (\ref{per2}), we see that once
the interferometer is biased, there appear terms in the
circulating current resulting from asymmetries in the couplings.
Collecting the results above, the circulating current, $I_{\rm
cir}$, can be written as
\begin{eqnarray}
I_{\rm cir}&=&I_{pc}+I_{a}.
\end{eqnarray}
Here, $I_{pc}$ denotes the part of the circulating current which
survives even when the system is un-biased, and is therefore
termed `persistent current',
\begin{eqnarray}
I_{pc}&=&e\int\frac{d\omega}{i\pi}\frac{f_{\ell}+f_{r}}{4}\Bigl
[\frac{\partial\Sigma^{R}_{\rm
ext}}{\partial\Phi}G^{R}_{dd}-cc\Bigr ].\label{IpcA}
\end{eqnarray}
The additional circulating current, that arises only when the
system is biased and there are asymmetries in the couplings is
denoted $I_{a}$,
\begin{eqnarray}
&&I_{a}=e i_{\ell}i_{r}j_{\ell}j_{r}(i\sin\Phi
)\int\frac{d\omega}{4\pi} (f_{r}-f_{\ell})\Delta \nonumber\\
&&\times\Bigl [(i_{r}^{2}-i_{\ell}^{2})|\alpha^{R}D^{R}_{0}|^{2}
\Bigl (G^{R}_{dd}+G^{R}_{dd}\nonumber\\
&+&(\Sigma^{R}_{\rm ext}+\Sigma^{A}_{\rm ext})
\frac{G^{R}_{dd}-G^{A}_{dd}}{\Sigma^{R}_{\rm ext}-\Sigma^{A}_{\rm
ext}}\Bigr )
+2(j_{r}^{2}-j_{\ell}^{2})|\alpha^{R}|^{2}\nonumber\\
&&\times D^{A}_{0}(1+\alpha^{R}D^{R}_{0}(i_{\ell}^{2}+i_{r}^{2}))
 \frac{G^{R}_{dd}-G^{A}_{dd}}{\Sigma^{R}_{\rm
ext}-\Sigma^{A}_{\rm ext}}\Bigr ].\label{IaA}
\end{eqnarray}
In particular, when the system is free of interactions, we use
Eqs. (\ref{g00}), and the analogous expression for $D^{R}_{0}$,
\begin{eqnarray}
\omega -\epsilon_{0}\equiv ({\cal D}^{0})^{-1}\equiv
(D^{R}_{0})^{-1}+\alpha^{R}(i_{\ell}^{2}+i_{r}^{2}),
\end{eqnarray}
to find that the `asymmetric' part of the interaction-free
circulating current is given by
\begin{eqnarray}
I_{a}^{0}&=&e i_{\ell}i_{r}j_{\ell}j_{r}(2i\sin\Phi )
\int\frac{d\omega}{4\pi}(f_{r}-f_{\ell})\Delta
|\alpha^{R}D^{R}_{0}G^{R0}_{dd}|^{2}\nonumber\\
&&\times\Bigl [ (i_{r}^{2}-i_{\ell}^{2})({\cal
G}^{0})^{-1}+(j_{r}^{2}-j_{\ell}^{2})({\cal D}^{0})^{-1}\Bigr ].
\end{eqnarray}

In the main text we omit this part of the circulating current,
that arises from the coupling asymmetries, and consider only the
term $I_{pc}$. Moreover, when the potential difference across the
interferometer is small (namely, the system is in the linear
response regime), one may neglect this difference altogether in
the sum $f_{\ell}+f_{r}$, and replace the electron distributions
by the thermal distribution one, Eq.  (\ref{thermal}). Note that
then, the relation (\ref{GddK}) becomes {\it exact}, and therefore
the result (\ref{IpcA}) does not rely on the wide-band
approximation. This is quite fortunate, since the persistent
current, as opposed to the transport current, requires integration
over the entire band. Hence, using for it an approximation which
is valid at a narrow range around the common Fermi energy is not
easily justifiable.

\end{multicols}
\end{document}